\documentclass[reprint,twocolumn,aps,prb,superscriptaddress]{revtex4-2}
\usepackage[T1]{fontenc}
\usepackage[utf8]{inputenc}
\usepackage{lmodern}
\usepackage{graphicx}
\usepackage{amsmath}
\usepackage{xcolor}
\usepackage{float}
\usepackage{hyperref}
\usepackage{braket}
\usepackage{amssymb,amsmath,amsthm, array}
\usepackage{mdframed}
\usepackage{systeme,mathtools}
\usepackage{yfonts}
\usepackage{comment}

\def  \bsigma   {\mbox{\boldmath$\sigma $}}
\def  \bnabla   {\mbox{\boldmath$\nabla $}}

\DeclareUnicodeCharacter{0301}{-}
\DeclareMathOperator\erf{erf}

\begin{document}
\renewcommand{\vec}[1]{\mathbf{#1}}
\newcommand{\ii}{\mathrm{i}}
\def\ya#1{{\color{orange}{#1}}}

\title{Topological insulator and quantum memory}

\author{M. Kulig}
\affiliation{Department of Physics and Medical Engineering, Rzesz\'ow University of Technology, 35-959 Rzesz\'ow, Poland}
\author{P. Kurashvili}
\affiliation{National Centre for Nuclear Research, Warsaw 00-681, Poland}
\author{C. Jasiukiewicz}
\affiliation{Department of Physics and Medical Engineering, Rzesz\'ow University of Technology, 35-959 Rzesz\'ow, Poland}
\author{M. Inglot}
\affiliation{Department of Physics and Medical Engineering, Rzesz\'ow University of Technology, 35-959 Rzesz\'ow, Poland}
\author{S. Wolski}
\affiliation{Department of Physics and Medical Engineering, Rzesz\'ow University of Technology, 35-959 Rzesz\'ow, Poland}
\author{S. Stagraczyński}
\affiliation{Institute of Spintronics and Quantum Information, Faculty of Physics, Adam Mickiewicz University, 61-614 Poznań, Poland}
\author{T. Masłowski}
\affiliation{Department of Physics and Medical Engineering, Rzesz\'ow University of Technology, 35-959 Rzesz\'ow, Poland}
\author{T. Szczepański}
\affiliation{Department of Physics and Medical Engineering, Rzesz\'ow University of Technology, 35-959 Rzesz\'ow, Poland}
\author{R. Stagraczyński}
\affiliation{Department of Physics and Medical Engineering, Rzesz\'ow University of Technology, 35-959 Rzesz\'ow, Poland}
\author{V. K. Dugaev}
\affiliation{Department of Physics and Medical Engineering, Rzesz\'ow University of Technology, 35-959 Rzesz\'ow, Poland}
\author{L. Chotorlishvili}
\affiliation{Department of Physics and Medical Engineering, Rzesz\'ow University of Technology, 35-959 Rzesz\'ow, Poland}

\date{\today}
\begin{abstract}
Measurements done on the quantum systems are too specific. Contrary to their classical counterparts, quantum measurements can be invasive and destroy the state of interest. Besides, quantumness limits the accuracy of measurements done on quantum systems. Uncertainty relations define the universal accuracy limit of the quantum measurements. Relatively recently, it was discovered that quantum correlations and quantum memory might reduce the uncertainty of quantum measurements. In the present work, we study two different types of measurements done on the topological system. Namely, we discuss measurements done on the spin operators and the canonical pair of operators: momentum and coordinate. We quantify the spin operator's measurements through the entropic measures of uncertainty and exploit the concept of quantum memory. While for the momentum and coordinate operators, we exploit the improved uncertainty relations. We discovered that quantum memory reduces the uncertainties of spin measurements. On the hand, we proved that the uncertainties in the measurements of the coordinate and momentum operators depend on the value of the momentum and are substantially enhanced at small distances between itinerant and localized electrons (the large momentum limit). We note that the topological nature of the system leads to the spin-momentum locking. The momentum of the electron depends on the spin and vice versa. Therefore, we suggest the indirect measurement scheme for the momentum and coordinate operators through the spin operator. Due to the factor of quantum memory, such indirect measurements in topological insulators have smaller uncertainties rather than direct measurements. 
\end{abstract}

\maketitle

\section{Introduction}
\label{sec:Introduction}

Heisenberg's uncertainty principle limits the precision of measurements done on momentum $\hat p$ and coordinate $\hat x$ of a quantum particle, i.e., pin-point measurement of one of the variables enhances uncertainty about the second variable and vice versa $\Delta x\Delta p\geqslant \hbar$, where $\hbar$ is  Planck's constant \cite{messiah2014quantum}. Note that from the point of view of classical mechanics, momentum and coordinate are a pair of canonical variables. From the point of view of quantum mechanics, momentum and coordinate operators violate commutativity $\left[\hat p,\hat x\right]\neq 0 $, i.e., the property of two operators considered together. Commutativity applies to a broader class of hermitian operators rather than canonically conjugate operators only. The generalization of Heisenberg's uncertainty principle done by Robertson applies to arbitrary two operators $\hat A,\,\hat B$ and has a form \cite{RevModPhys.89.015002}: 
\begin{eqnarray}\label{Robertson}
\Delta A\cdot\Delta B\geqslant \frac{1}{2}\left\vert\bra{\psi}[\hat A,\hat B]\ket{\psi}\right\vert, 
\end{eqnarray}
where $\Delta O=\sqrt{\langle \hat O^2\rangle-\langle\hat O\rangle^2}$. 
While Eq.(\ref{Robertson}) is more general than Heisenberg's relation, it still depends on the choice of the state over which averaging is done. Consequently, when the state  $\ket{\psi}$ is the eigenfunction of one of the operators $\hat A$ or $\hat B$, Eq.(\ref{Robertson}) takes a trivial form. Maccone and Pati tried to avoid this problem and quantified uncertainty through two orthogonal states as follows
\cite{PhysRevLett.113.260401}:
\begin{eqnarray}\label{Maccone and Pati}
&&\Delta A^2+\Delta B^2\geqslant i\bra{\psi}[\hat A,\hat B]\ket{\psi}+\nonumber\\
&&\vert\bra{\psi}\hat A+i\hat B\ket{\psi^\bot}\vert.
\end{eqnarray}
Here $\ket{\psi^\bot}$ is the state orthogonal to $\ket{\psi}$. 
The early attempts at studying quantum uncertainty relations concerned a proper choice of quantum states. However, remarkable recent progress was achieved through reforming the uncertainty paradigm to the entropic measures and entropic uncertainty relations. We admit a celebrated work \cite{berta2010uncertainty}. The core concepts of entropic uncertainty relations are viable for quantum guessing games and quantum memory \cite{wang2019quantum,ming2020improved,dolatkhah2020tightening,bergh2021entanglement, PhysRevA.104.062204,chotorlishvili2019spin,song2022environment,zhu2021zero,kurashvili2022quantum, PhysRevD.103.036011,kurashvili2022quantum}.

In what follows, we study a guessing game between two parties Alice (\textbf{A}) and Bob (\textbf{B}). We implement tools of quantum metrology to the experimentally feasible condensed matter system: a quantum dot placed on the surface of a topological insulator (TI). The model of interest is described in the next section in detail. Here we specify the rules of the quantum game that has to be played. In the quantum game, Bob represents a quantum dot (QD) placed on the surface of a topological insulator (TI). Bob selects a single electron \textbf{A} from the surface of the TI and scatters electron \textbf{A} on the electron  \textbf{B} localized in the QD. The scattered electron \textbf{A} Bob shares with Alice and lets her perform two consecutive measurements on $A$. Alice measures two $Z$ and $X$ components of the spin. A measurements done on \textbf{A} Alice denotes by $R\equiv \lbrace Z, X\rbrace$ and stores measurements results in \textbf{L}. The aim of Bob is to guess the results of the measurements $R\equiv \lbrace Z, X\rbrace$. Bob's uncertainty about the measurements result can be reduced by quantum memory.  
In the present project, we analyze two types of measurements done on the system: measurements done on the momentum and coordination of the itinerant electron and spin projections of the itinerant electron. To explore uncertainties of the measurements done on the momentum and coordinate, we follow Maccone and Pati \cite{PhysRevLett.113.260401}. When quantifying uncertainties of the spin measurements, we exploit the quantum memory. In the first case, we show that uncertainty increases with the momentum $k$ (i.e, at the shorter distances between localized and itinerant electrons). On the other hand, the spin measurements' uncertainty is reduced by the quantum memory and is independent of $k$. The paper is organized as follows: in section \textbf{II}, we describe the model. In section \textbf{III}, describe uncertainties of the momentum and coordinate measurements. In section \textbf{IV}, we analyze entropic measures of spin measurements and quantum memory. In section V we conclude the work. 

\section{Model}
\label{sec:Model}

The system of our interest consists of a 2D surface itinerant electron in a topological insulator  $\hat H_{I}$, localized electron in a quantum dot placed on the surface of topological insulator  $\hat H_{D}$ and exchange interaction $\hat V$ between itinerant and localized electrons:
\begin{eqnarray}
\label{The total Hamiltonian of the system}
&& \hat H_{tot}=\hat H_{I}+\hat H_{D}+\hat V.
\end{eqnarray}
The Hamiltonian of the surface electron explicitly reads \cite{PhysRevLett.105.266806}:  $\hat H_{I}=-iv\, \hat{\bsigma }_A\cdot \bnabla $, where $\hat{\bsigma }_A$ is a Pauli matrix
and $v$ is the velocity of the surface electron. The eigenstates and eigenenergies of the surface electrons can be found easily $\psi_{T,\sigma}(\textbf{r})
=\frac{e^{i\textbf{k}\cdot \textbf{r}}}{\sqrt{2}} \begin{pmatrix} 1 \\ \pm k_+/k \end{pmatrix}$, $E =\pm vk$,
where $k_+=k_x+ik_y$, $k=\sqrt{k_x^2+k_y^2}$ and $\textbf{r}=(x,y)$. The Hamiltonian of electron localized in the quantum dot (QD) has the form:
\begin{eqnarray}\label{Hamiltonian of the electron localized}
&&\hat H_{D}=-B\hat\sigma_B^z-\frac{\hbar^2}{2m}\vec\nabla^2+\frac{1}{2}m\omega_0^2r^2,
\end{eqnarray}
where $\omega_0$ is the frequency of electron oscillation in QD.  
Through the external magnetic field applied locally to the quantum dot (e.g., through spin-polarized scanning tunneling microscopy (SP-STM) \cite{PhysRevResearch.3.043185,willke2019tuning}), we can freeze (strong field) or relax (weak field) the spin of the localized electron $\hat\sigma_B^z$ depending on the value of Zeeman splitting $B\equiv \hbar\gamma_eB$.
The lowest eigenstate of the localized electron has the form 
\begin{eqnarray}\label{confined in the quantum dot}
&&\psi_{D}(\textbf{r})=\frac{1}{l_B\sqrt{\pi}}\, \exp\left(-\frac{x^2+y^2}{2l_B^2}\right) ,
\nonumber\\
&& l_B^2=l_0^2/\sqrt{1+B^2e^2l_0^4/4\hbar^2}.
\end{eqnarray}
Here $l_0=(\hbar/m\omega_0)^{1/2}$ is the confinement length \cite{PhysRevB.82.045311}. The last term in Eq.(\ref{The total Hamiltonian of the system}) describes the interaction between localized and surface electrons $\hat V =J\, \hat\bsigma _A\cdot \hat\bsigma _B\; \delta\left(\textbf{r}_1-\textbf{r}_2\right)$. The origin of the exchange interaction and exchange constant $J$ are described in \cite{PhysRevB.89.075426}. In what follows $E\equiv\langle E\rangle=\langle vk\rangle$ and  set the dimensionless parameters through $\hat H_{tot}\to\hat H_{tot}/J$. 

Following the recent work \cite{PhysRevB.106.224418} we consider the spin-dependent formulation of the scattering problem and exploit the Lippmann–Schwinger integral equation. 
The initial wave function is a product state $\psi_{T,\sigma}(\textbf{r}_A)=\psi(\textbf{r}_A)\left(\alpha\vert 0\rangle_A+\beta\vert 1\rangle_A\right)$ and $\psi_{D,\sigma}(\textbf{r}_B)=\psi(\textbf{r}_B)\vert 0\rangle_B$, and for brevity, in what follows $\textbf{r}_A\equiv\textbf{r}_1$ and  $\textbf{r}_B\equiv\textbf{r}_2$.  Two states of localized electron $\psi_{D,0}(\textbf{r})=\psi_D(\textbf{r})\vert 0\rangle$ (spin-up, $|0\rangle\equiv\ket{\uparrow}$) and  $\psi_{D,1}(\textbf{r})=\psi_D(\textbf{r})\vert 1\rangle$ (spin-down, $|1\rangle\equiv\ket{\downarrow}$) with the respective energies $\varepsilon_0$ and $\varepsilon_1=\varepsilon_0+2B$ (hereafter, we set $\varepsilon_0=0$) contribute to the post-scattering state. The post-scattering wave function of the two-electron system reads \cite{PhysRevB.106.224418}: $\Psi_{\sigma_1\sigma_2}(\textbf{r}_1,\textbf{r}_2)=\psi_{T,0}^{(+)}(\textbf{r}_1)\psi_{D,0}(\textbf{r}_2)+\psi_{T,1}^{(+)}(\textbf{r}_1)\psi_{D,1}(\textbf{r}_2)$, where spinors $\psi_{T,0}^{(+)}(\textbf{r})= \begin{pmatrix} \phi_0(\textbf{r}) \\  \chi_0(\textbf{r}) \end{pmatrix}$ and $\psi_{T,1}^{(+)}(\textbf{r})= \begin{pmatrix} \phi_1(\textbf{r}) \\  \chi_1(\textbf{r})
\end{pmatrix}$ are found from the coupled integral equations 
\begin{widetext}
\begin{align}
\begin{pmatrix} \phi_0(\textbf{r}) \\ \chi_0(\textbf{r}) \end{pmatrix} &=\frac{e^{i\textbf{kr}}}{\sqrt{2}}
\begin{pmatrix} 1 \\  k_+/k \end{pmatrix} +\int d^2r'\,\hat{G}^{(+)}(\textbf{r},\textbf{r}';E)
\hat{V}_{00}(\textbf{r}')\begin{pmatrix} \phi_0(\textbf{r}') \\  \chi_0(\textbf{r}') \end{pmatrix}
+\int d^2r'\,\hat{G}^{(+)}(\textbf{r},\textbf{r}';E)
\hat{V}_{01}(\textbf{r}')\begin{pmatrix} \phi_1(\textbf{r}') \\  \chi_1(\textbf{r}') \end{pmatrix},\nonumber
\end{align}
\begin{align}\label{coupled integral equations second}
\begin{pmatrix} \phi_1(\textbf{r}) \\ \chi_1(\textbf{r}) \end{pmatrix} &=\int d^2r'\,\hat{G}^{(+)}(\textbf{r},\textbf{r}';E-2B)\hat{V}_{10}(\textbf{r}')\begin{pmatrix} \phi_0(\textbf{r}') \\  \chi_0(\textbf{r}') \end{pmatrix}+\int d^2r'\,\hat{G}^{(+)}(\textbf{r},\textbf{r}';E-2B)
\hat{V}_{11}(\textbf{r}')\begin{pmatrix} \phi_1(\textbf{r}') \\  \chi_1(\textbf{r}') \end{pmatrix},
\end{align}
\end{widetext}
where $E=vk$ is the energy of the itinerant electron and the Green's function is given by $\hat{G}^{(+)}(\textbf{r},\textbf{r}';E)=-\frac{E}{2\pi v^2}[K_0(-iE|\textbf{r}-\textbf{r}'|/v)\hat{I}+
K_1(-iE|\textbf{r}-\textbf{r}'|/v)\hat{\sigma}^x]$, and $K_{0,1}$ are the modified Bessel functions (the Macdonald functions). We skip cumbersome details \cite{PhysRevB.106.224418} of the solution of Eq.(\ref{coupled integral equations second}) and present the final result: 
\begin{eqnarray}\label{after cumbersome calculations}
&&\ket{\psi}=C_1\ket{0}_A\ket{0}_B+C_2\ket{0}_A\ket{1}_B+\nonumber\\
&&C_3\ket{1}_A\ket{0}_B+C_4\ket{1}_A\ket{1}_B. 
\end{eqnarray}
Here we introduced the following notations:
\begin{widetext}
\begin{eqnarray}\label{here one}
&& C_1\left(\rho,\varphi\right)=\frac{\psi_D(\rho')}{2\pi\sqrt{2}C_0} 
\left(2\pi e^{ik\rho\cos\varphi}+k_Jk\left({A}_{00}(\rho,\varphi)+e^{i\theta}{A}_{10}(\rho,\varphi)\right)\right),\nonumber\\
&&C_2\left(\rho,\varphi\right)=-\frac{\psi_D(\rho')}{2\pi\sqrt{2}C_0} 
\left(2k_J(k-2k_B){A}_{0B}(\rho,\varphi)\right)e^{i\theta},\nonumber\\
&&C_3\left(\rho,\varphi\right)=\frac{\psi_D(\rho')}{2\pi\sqrt{2}C_0}\left(2\pi e^{ik\rho\cos(\varphi)}+k_Jk\left({A}_{00}(\rho,\varphi)+e^{-i\theta}{A}_{10}(\rho,\varphi)\right)\right)e^{i\theta},\nonumber\\
&&C_4\left(\rho,\varphi\right)=-\frac{\psi_D(\rho')}{2\pi\sqrt{2}C_0} 
\left(2k_J(k-2k_B){A}_{1B}(\rho,\varphi)\right)e^{i\theta},\nonumber\\
&&C_0=\sqrt{\vert C_1\vert^2+\vert C_2\vert^2+\vert C_3\vert^2+\vert C_4\vert^2},\,\,\,
\end{eqnarray}
\end{widetext}
where $\varrho$, $\varphi$ are polar coordinates and coefficients $\textbf{A}$ are defined in the Appendix. Dimensionless parameters are introduced through the following notations $k_+/k=e^{-i\theta}$, $J/v=k_J$, $E\equiv E/J$, $E_B=E-2B$, $E/v=k$, $E_B/v=k-2k_B$, $k_B\equiv B/v$, and $l=0,1$.
In the coefficients $A_{lB}(\mathbf{r})$, indices $l=0,1$ define zero and the first-order modified Bessel functions and $B=0,1$ indicates on the zero and nonzero magnetic field problems.

\section{Momentum and coordinate uncertainties}
\label{sec: Momentum and coordinate}

Exploiting the improved uncertainty relations by Maccone and Pati,  Eq.(\ref{Maccone and Pati}) in polar coordinates is not a trivial question. Following \cite{messiah2014quantum} and \cite{khelashvili2022generalized}, we define the operators for TI and QD as follows: $\hat p_\rho=-i\hbar\left(\frac{d}{d\rho}+\frac{1}{\rho}\right)$ and $\hat p_{\rho'}=-i\hbar\left(\frac{d}{d\rho'}+\frac{1}{\rho'}\right)$ and obtain commutation relations $[\rho, \hat p_\rho]=i\hbar$,
$[\rho', \hat p_{\rho'}]=i\hbar$, $[\rho^2, \hat p^2_\rho]=4\hbar^2\left(\rho\frac{d}{d\rho}+\frac{3}{2}\right)$,
$[\rho'^2, \hat p^2_{\rho'}]=4\hbar^2\left(\rho'\frac{d}{d\rho'}+\frac{3}{2}\right)$. 
The operator $\hat p_\varphi=-i\frac{\partial}{\partial\varphi}$ leads to the complications. It is Hermitian in the space of functions with the period $2\pi$. However, if we restrict discussion to this range, then commutation relations should be modified following the work \cite{judge1963commutator}, meaning that one has to include a series of $\delta$ functions:
\begin{eqnarray}\label{include a series}
[\hat p_\varphi, \varphi]=i\left\lbrace1-2\pi\sum\limits_{n=-\infty}^{+\infty}\delta[\varphi-(2n+1)\pi]\right\rbrace. 
\end{eqnarray}
In order to avoid complications with Eq.(\ref{include a series}) we consider two cases \cite{PhysRevA.22.797,chotorlishvili2017zitterbewegung} commutator 
$[\sin\varphi,\hat p_\varphi]=i\cos\varphi$ and consider the operators, $\hat S=e^{i\varphi}$, $\hat P=\hat S\frac{\partial}{\partial \hat S}$, $[\hat P,\hat S]=\hat S$. 
Taking into account Eq.(\ref{Maccone and Pati}) and 
Eq.(\ref{after cumbersome calculations}), 
Eq.(\ref{here one}) after cumbersome calculations we deduce:

\begin{eqnarray}\label{improved Momentum and coordinate}
&&\Delta P^2+\Delta S^2\geqslant i\bra{\psi}[\hat P,\hat S]\ket{\psi}+\nonumber\\
&&\vert\bra{\psi}\hat P+i\hat S\ket{\psi^\bot}\vert,\nonumber\\
&&\Delta (\sin\varphi)^2+\Delta p_\varphi^2\geqslant i\bra{\psi}[\sin\varphi,\hat p_\varphi]\ket{\psi}+\nonumber\\
&&\vert\bra{\psi}\sin\varphi+i\hat p_\varphi\ket{\psi^\bot}\vert,\\
&&\Delta p_\rho^2+\Delta \rho^2\geqslant i\bra{\psi}[\hat p_\rho,\hat \rho]\ket{\psi}+\nonumber\\
&&\vert\bra{\psi}\hat p_\rho+i\hat \rho\ket{\psi^\bot}\vert\nonumber.
\end{eqnarray}

Applying Eq.(\ref{improved Momentum and coordinate}) to realistic physical systems of interest technically is rather demanding. Therefore we calculate only one particular relation (hereafter $\hbar=1$):
\begin{widetext} 
\begin{eqnarray}\label{one particular case}
\Delta p_\rho^2+\Delta \rho^2\geqslant 1+
\vert\bra{\psi}\rho-\left(\frac{d}{d\rho}+\frac{1}{\rho}\right)\ket{\psi^\bot}\vert.
\label{eq:uncertainty_rho}
\end{eqnarray}
\end{widetext}
The set of the states $\mathbb{C}=\lbrace \Psi^\perp\rbrace$ orthogonal to  $\ket{\psi}$ we express in terms of the coefficients Eq.(\ref{here one}) as follows
\begin{eqnarray}
\label{eq:Psi_perp1}
&& \Psi^\perp_1 = \{C_2, -C_1, C_4, -C_3\},
\\
\label{eq:Psi_perp2}
&& \Psi^\perp_2 = \{C_3, -C_4, -C_1, C_2\},
\\
\label{eq:Psi_perp3}
&& \Psi^\perp_3 = \{C_4, -C_3, C_2, -C_1\}.
\end{eqnarray}
The general orthogonal wave function reads
\begin{equation}
\ket{\psi^\perp} = \sum_{i=b}^3 \alpha_b \ket{{\Psi^\perp}_b},
\label{eq:psi_perp}
\end{equation}
where $\vert\alpha_1\vert^2+\vert\alpha_2\vert^2 + \vert\alpha_3\vert^2=1$.
Then we rewrite the general orthogonal function in the computational basis 
Eq.(\ref{after cumbersome calculations})
\begin{equation}
\psi^\perp = \sum C_i^\perp \ket{\psi}_{AB}.
\label{eq:psi_perp}
\end{equation}
The coefficients $C_i^\perp$ are expressed through the coefficients
$\{C_i\}$ and parameters $\{\alpha\}$:
\begin{eqnarray}
&& C_1^\perp = \alpha_1 C_2 + \alpha_2 C_3 +\alpha_3 C_4,
\label{eq:C1_perp}
\\
&& C_2^\perp = -\alpha_1 C_1 -\alpha_2 C_4 - \alpha_3 C_3,
\label{eq:C2_perp}
\\
&& C_3^\perp = \alpha_1 C_4 - \alpha_2 C_1 +\alpha_3 C_2,
\label{eq:C3_perp}
\\
&& C_4^\perp = -\alpha_1 C_3 + \alpha_2 C_2 - \alpha_3 C_1.
\label{eq:C4_perp}
\end{eqnarray}
Let $\hat R$ be an arbitrary operator and 
$R_{ik} =\int C^*_i\hat R C_k\rho d\rho d\varphi$, it's matrix element. Then
\begin{eqnarray}
\nonumber
&& \bra{\psi}R \ket{\psi^\perp} = 
\\
\nonumber
&& \quad \alpha_1 [R_{12}-R_{21}+R_{34}-R_{43}] +
\\
\nonumber
&& \quad \alpha_2[R_{13}-R_{31} -R_{24} +R_{42}]+
\\
&& \quad \alpha_3[R_{14}-R_{41} -R_{23}+R_{32}].
\label{eq:R_matrix_element}
\end{eqnarray}
Comparing Eq.(\ref{eq:R_matrix_element}) to Eq. (\ref{eq:uncertainty_rho}), we can see, that
the right hand side of Eq. (\ref{eq:uncertainty_rho}) is larger than one only if $R_{ik}\neq R_{ki}$.
Therefore the nonzero $\bra{\psi}\rho-(d_\rho+1/\rho)\ket{\psi}^\perp$ is granted by non-hermitian part in the operator $\hat R$.
As it is shown in Appendix, the functions $A_{ib}$ in the expressions for the
coefficients $C_i(\rho, \varphi)$ can be written as a product of functions 
of $\rho$ and $\varphi$:
\begin{equation}
\label{eq:expression_for_A}
A_{lb}(\rho, \varphi) =  f_{\lambda b}(\rho) g_b(\varphi),
\end{equation}
The first index, $l=0,1$ arises from the expansion of the Bessel-Macdonald
functions.
In the following we use $\lambda = 1-4l^2$, which is equal to either
$1$ or $-3$.
The parameter of the angular function, $b$, takes the values $0$ and $B$ 
for zero and non-zero magnetic field respectively.
We set $l_B=1$ and obtain:
\begin{eqnarray}
&& f_{b\lambda}(\rho) = \frac {\pi }{8k_{b}^2} e^{ik_b\rho}
\left(\frac{\lambda}{\rho^3} + \frac{8 i k_b }{\rho^2}\right),
\label{eq:f_rhof}
\\
&& g_b(\varphi) = a_{0b} + a_{1b} \cos \varphi + a_{2b} \cos^2 \varphi,
\label{eq:g_varphi}
\end{eqnarray}
where $\lambda = 1-4l^2$, 
$a_{0b}=1- (k_{(b)}^2/2)-(i\sqrt \pi/ 2) k_{(b)}$,
$a_{1b}=[ k_{(b)}+ i\sqrt \pi/2]k $,
$a_{2b}= k^2 /2$,
and $k_b$ is either $k$ or $k - 2k_B$ depending on the absence or presence of the magnetic
field.

The integrals of the uncertainty relation have the form
\begin{eqnarray}
\nonumber 
&& \int \rho d \rho d\varphi A^*_{l_1b_1}(\rho, \varphi)
\frac \partial {\partial \rho} A_{l_2b_2}(\rho, \varphi) = 
\\
\nonumber 
&& \qquad \int d\varphi g_{b_1}(\varphi) g_{b_2}(\varphi) 
\int \rho d\rho f_{\lambda_1b_1}(\rho)\frac d {d\rho}f_{\lambda_2b_2}(\rho)=
\\
&&  \qquad \qquad \qquad \qquad \qquad 
I^\varphi_{b_1b_2} I^R_{\lambda_1\lambda_2,b_1b_2}.
\label{eq:integrals_UR}
\end{eqnarray}
The integration over the angular variable is rather straightforward and yields
\begin{eqnarray}
\nonumber
&& I^\varphi_{b_1b_2}=\int_{0}^{2\pi}d\varphi g_{b1}^*(\varphi)g_{b2}(\varphi)= 
2\pi a^*_{0b_1}a_{0b_2} +
\\
\nonumber 
&& \qquad \pi(a^*_{1b_1}a_{1b_2} + a^*_{0b_1}a_{2b_1} + a^*_{2b_1}a_{0b_2})
+ 
\\
&& \qquad \qquad \qquad 
\frac {3\pi}{4} a^*_{2b_2}a_{2b_2}.
\label{eq:Ivarphi}
\end{eqnarray}
In Appendix \ref{sec:expressions_for_integrals} we write explicitly the form
of these integrals for all possible values of the magnetic field.
The result of the radial integrations reads:
\begin{eqnarray}
\nonumber
&& I^R_{\lambda_1\lambda_2 b_1b_2} = 
\int_{1}^{\infty}\rho d\rho f^*_{\lambda_1b_1}(\rho)
\frac d {d\rho} f_{\lambda_2b_2}(\rho)=
\\
&&\quad \left[ \frac{\pi}{k_{b1}k_{b2}} \right]^2
\sum_{n=2}^{5}F_ni^{n} (-i\Delta k_{b})^n \Gamma(-n, -i \Delta k_{b}),
\label{eq:Ir}
\end{eqnarray}
where $k_{1,2b}= k-2k_{b_{1,2}} = (E-2b_{1,2})/v$, 
$\Delta k_{(b)} = k_{1b} - k_{2b} = 2(b_2-b_1)/v$, $b_{1,2}=0, B$, and
the coefficients $F_n$ are
\begin{eqnarray}
&& F_2 = k_{1b}k_{2b}^2,
\label{eq:F2}
\\
&& F_3 = \frac 1 8 \lambda_1 k_{2b}^2 - \frac 1 8 \lambda_2 k_{1b}k_{2b}
+2k_{1b}k_{2b},
\label{eq:F3}
\\
&& F_4 = \frac 1 {64} \lambda_1\lambda_2k_{2b}
\label{eq:F4}
\\
&& F_5 = \frac {\lambda_1\lambda_2} {32},
\label{eq:F5}
\end{eqnarray}
$\lambda_{0,1}=1-4l_{0,1}^2$ can take on values either $1$ ($l=0$) or $-3$
($l=1$).

The incomplete gamma-functions with a negative integer index, $\Gamma(-n,x)$,
satisfy the following condition in the limit $x\rightarrow\infty$
\begin{equation}
\lim_{x\rightarrow 0} x^n \Gamma(-n, x) = \frac 1 n.
\label{eq:GammaN_at_zero}
\end{equation}
In Appendix \ref{sec:normalization}, we also calculate the sum of squares 
of the absolute values of the coefficients $C_i$ truncating the contribution from the plane waves. We proceed to the uncertainty relation in  Eq. (\ref{eq:R_matrix_element}). 
The simplest case corresponds to $\alpha_1=0$, $\alpha_2 =1$ ,$\alpha_3=0$.
Then, we calculate the differences between matrix elements 
$R_{13} - R_{31}$ and $R_{24}-R_{42}$, where $\hat R$ corresponds to 
the differentiation by $\rho$.
Then antisymmetric combinations of the  matrix elements read:
\begin{eqnarray}
\nonumber 
&& R_{13}-R_{31} = \int \rho d\rho d\varphi \left[C^*_1\frac d {d\rho} C_3
- C^*_3\frac d {d\rho}C_1\right]=
\\
&& -2 iA_{13}
\left(I^R_{11,00}\sin2\theta+\sin\theta[I^R_{10,00}+I^R_{01,00}]\right)
I^\varphi_{00},
\label{eq:R13}
\\
&& R_{24}-R_{42} = 4 A_{24}\left( I^R_{01,BB}-I^R_{10,BB}\right)I^\varphi_{BB},
\label{eq:R24}
\end{eqnarray}
where
\begin{eqnarray}
&& A_{13} = \frac{\vert\psi_D(\rho')\vert^2 k_J^2k^2}{8\pi^2\vert C_0\vert^2},
\label{eq:A13}
\\
&& A_{24} = \frac{\vert\psi_D(\rho')\vert^2 k_J^2(k-2k_B)^2}
{8\pi^2\vert C_0\vert^2}.
\label{eq:A24}
\end{eqnarray}
We write down the integrals from Eqs. (\ref{eq:R13}) and (\ref{eq:R24}).
The results of the angular integration $I^\varphi_{00}$,
$I^\varphi_{BB}$, are presented in Eqs. (\ref{eq:I00}) and (\ref{eq:IBB}) of
Appendix \ref{sec:expressions_for_integrals}.
For radial integrals, we have:
\begin{eqnarray}
&& I^R_{11,00} = -\frac {\pi^2}{k^2}\left[ \frac k 2 (k^2 - \frac 1 {128})
+i(\frac 2 3 k^2 - \frac 1 {160})\right],
\label{eq:IR1100}
\\
\nonumber
&&  I^R_{10,00} + I^R_{01,00} = - \frac {\pi^2} {k^2} \times
\\
&& \qquad \qquad 
\left [ k(k^2+\frac 1 {128}) +i (\frac{k^2} 3 +\frac 3 {80})\right],
\label{eq:sum_IR1000_IR0100}
\\
&&  I^R_{10,BB} - I^R_{01,BB} = 0.
\label{eq:diff_IR10BB_IR01_BB}
\end{eqnarray}
We see that for the transverse wave function our choice,
only the term $R_{13}-R_{31}$ given by Eq. (\ref{eq:R13}) contributes to the uncertainty relation. Summarizing the obtained results, the entire expression of the uncertainty reads
\begin{eqnarray}
\nonumber
&& \Delta p_\rho^2 + \Delta \rho^2 \geq 1 + \vert \bra{\psi} \frac d {d\rho} 
\ket{\psi^\perp} \vert =
\\
\nonumber 
&&\qquad  1+\bigg|C^*_1\frac d {d\rho}C_3-C^*_3\frac d {d\rho}C_1  +
\\
&& \qquad \qquad \qquad C^*_2\frac d {d\rho}C_4-C^*_4\frac d {d\rho}C_2 \bigg|.
\label{the entire expression of the uncertainty}
\end{eqnarray}
The coefficients $C_i$ are similar to those given in
Eq. (\ref{after cumbersome calculations}) without contribution from the plane wave components
(first terms inside the parenthesis in the expressions for $C_1$ and $C_3$).
In this case, $C_0$ is normalized by means of integration over the entire space, as it is discussed in Appendix \ref{sec:normalization}.
The coefficient $\vert C_0 \vert^2$  is equal to
\begin{eqnarray}
\nonumber
&&
\vert C_0 \vert^2 =\frac{\vert\psi_D(\rho')\vert^2} 4 \pi k_J^2\times
\\
\nonumber
&& \qquad \bigg[\frac{I^\varphi_{00}}{k^2} \left(k^2(1+\cos\theta) - \frac 1 {32}
(1+2\cos \theta)\right) + 
\\
&& \qquad \qquad \qquad \frac{I^\varphi_{BB}}{(k-2k_B)^2}
((k-2k_B)^2 - \frac 1 {32} ) \bigg].
\label{eq:C0_square}
\end{eqnarray}
The coefficients $I^\varphi_{00}$ and $I^\varphi_{BB}$ arise from the angular
integration of $A_{lb}(\rho, \varphi)$ and are written explicitly in
Eqs. (\ref{eq:I00}) and (\ref{eq:IBB}).
The integrals of the radial part are expressed in terms of incomplete
gamma-functions, and it can be proved that in Eq.(\ref{the entire expression of the uncertainty}) only first two terms $C^*_1\frac d {d\rho}C_3,\,C^*_3\frac d {d\rho}C_1 $
contribute to the uncertainty.

Without the normalization factors, those terms are equal to the sum of two terms proportional to $\sin 2\theta$ and $\sin \theta$, respectively, 
see Eq. (\ref{eq:R13}) and the values of the coefficients are given in Eqs. (\ref{eq:IR1100}) and (\ref{eq:sum_IR1000_IR0100}).
In this approximation, dependence on the magnetic field in the form of $k_B$ remains only in one of the normalization constants, $C_0$.
To express our results in graphical form, we consider the center of the localized electron $\rho'=0$, and $\vert\psi_D(\rho')\vert^2=1$.

\begin{figure}[h!]
\includegraphics[width=\columnwidth]{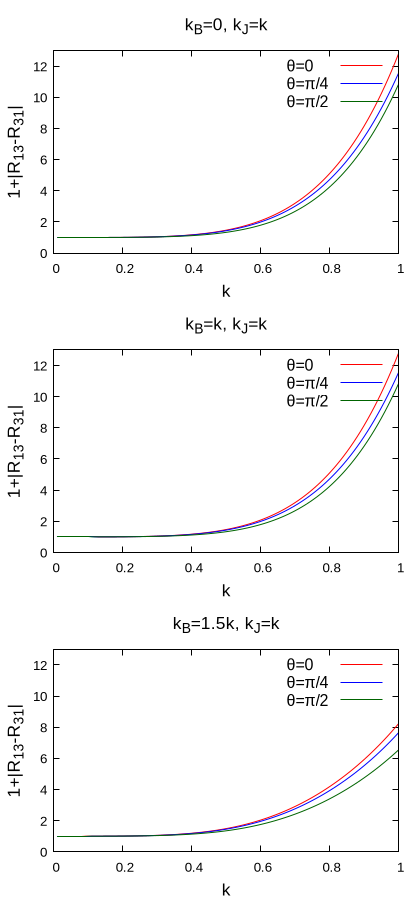}
\caption{
\label{fig:Rdiff}
Dependencies of the minimum value of the uncertainty relation as calculated
from Eq. (\ref{the entire expression of the uncertainty}) for different coefficient of proportionality between
$k_B$ and $k$ at fixed $k_J=k$ and three different values of $\theta$. 
}
\end{figure}
In Fig.\ref{fig:Rdiff}, we plot the right-hand side of the equation Eq.(\ref{the entire expression of the uncertainty}). The value above the one is related to the contribution from the term  $\big|C^*_1\frac d {d\rho}C_3-C^*_3\frac d {d\rho}C_1  +
\qquad \qquad \qquad C^*_2\frac d {d\rho}C_4-C^*_4\frac d {d\rho}C_2 \big|$. 
As we see, the uncertainty depends on the scattering angle and magnetic field. In the case of the strong magnetic field, $k_B=1,5k$ uncertainty is smaller. On the other hand, uncertainty increases with the distance. In what follows, we show that it is not the case for the quantum memory and entropic measures of the uncertainty. 

\section{Quantum guessing game}
\label{sec: guessing game}

 Before starting the quantum guessing game between Alice and Bob, sharing the system of TI and QD, we define quantities of interest and tools. Taking into account the wave function of the system $\ket{\psi}_{AB}$, we construct the density matrix $\hat\varrho_{AB}=\ket{\psi}\bra{\psi}_{AB}$. Then, the reduced density matrices of the TI and QD subsystems are defined as follows $\hat\varrho_A=Tr_B\left(\hat\varrho_{AB}\right)$ and $\hat\varrho_B=Tr_A\left(\hat\varrho_{AB}\right)$. The von Neumann Entropy of the entire system is given by  $S(AB)=-Tr\left(\hat\varrho_{AB}\log\hat\varrho_{AB}\right)$. The conditional quantum entropy has a form $S(A\vert B)=S(AB)-S(B)$. After Alice performs two measurements and measures $z$ and $x$ components of her qubit \textbf{A}, the subsequent post-measurement density matrices are given through
\begin{widetext}
\begin{eqnarray}
\label{eq:rhoZ}
&&\hat\varrho_{Z,AB}=\sum\limits_{n}\ket{\psi_n}\bra{\psi_n}_A\otimes Tr_A\left\lbrace(\ket{\psi_n}\bra{\psi_n}_A\otimes\hat I_B)\hat\varrho_{AB}\right\rbrace,\\
\label{eq:rhoX}
&&\hat\varrho_{X,AB}=\sum\limits_{n}\ket{\phi_n}\bra{\phi_n}_A\otimes Tr_A\left\lbrace(\ket{\phi_n}\bra{\phi_n}_A\otimes\hat I_B)\hat\varrho_{AB}\right\rbrace.
\end{eqnarray}
\end{widetext}
Here $\ket{\psi_{1,2}}\equiv \ket{0}_A,\,\ket{1}_A$ and $\ket{\phi_{1,2}}=\frac{1}{\sqrt{2}}\left(\ket{0}_A\pm\ket{1}_A\right)$ are eigen functions of the $z$, $x$ components of the qubit $\textbf{A}$.

Through the scattering process, Bob entangles the particle \textbf{B} QD with the particle \textbf{A} TI. Alice performs two measurements on her particle   \textbf{A} and  broadcasts her measurement choice to Bob. Bob wants to guess Alice's outcome precisely by measuring his particle B with the help of the received classical information (i.e., Alice's choice of measurement). Bob's ignorance about Alice's measurements is given by:

\begin{eqnarray}
\label{eq:BobsIgnorance}
&& S(X\vert B)+S(Z\vert B)\geqslant \log_2(1/c)+S(A\vert B),\\
\label{eq:cij}
&& c_{ij}=\max\left\lbrace \vert\bra{\psi_i}\ket{\phi_j}\vert^2\right\rbrace. 
\end{eqnarray}
$c$ is a measure of complimentarity. \vspace{0.5cm}\\
In essence, this scheme allows one to read out information about the quantum dot through the measurements done on the itinerant electron from the topological insulator or vice versa, depending on the experimental feasibility and convenience. 

The density operator of the bipartite system $\hat \varrho_{AB}$ we present in the form:
\begin{equation}
\label{eq:rhoABexpansion}
\hat \varrho_{AB} = \sum_{iklm} \rho_{iklm}
\ket{\psi^A_{i}\psi^B_{k}}\bra{\psi^A_{l}\psi^B_{m}}.
\end{equation}
The indices $\{iklm\}$ run over the components of the basis states 
of the spins belonging to Alice and Bob.
The matrix elements $\rho_{iklm}$ are obtained from the coefficients $C_i$ in 
Eq. (\ref{after cumbersome calculations}):
\begin{equation}
\label{eq:cFromRho}
\rho_{iklm} = C_{2i+k+1}C^*_{2l+m+1} = c_{ik}c^*_{lm},
\end{equation}
where $c_{00} = C_1$, $c_{01}=C_2$, $c_{10} =C_3$, $c_{11}=C_4$.
After inserting Eq. (\ref{eq:rhoABexpansion}) into Eqs. (\ref{eq:rhoZ}) and
(\ref{eq:rhoX}), we deduce:
\begin{eqnarray}
\label{eq:rhoZnew}
&&\hat \varrho_{Z,AB} = \sum_{ikm} \rho_{ikim}
\ket{\psi^A_i \psi^B_{k}}\bra{\psi^A_i\psi^B_{m}},
\\
&&\hat \varrho_{X,AB} = \sum_{ikm} \rho^{(\Phi)}_{ikim}
\ket{\phi^A_i \phi^B_{k}}\bra{\phi^A_i\phi^B_{m}},
\end{eqnarray}
where $\rho^{\Phi}_{\{i\}}$
are the matrix coefficients in the basis $\{\phi_1, \phi_2\}$,
In order to calculate $\rho^{\Phi}_{\{i\}}$ we exploited Eqs. (\ref{after cumbersome calculations}) 
and (\ref{eq:cFromRho}) and the coefficients
$C_i^{(\Phi)}$ in the $X$-basis.  The details of derivations are given in Appendix \ref{sec:app_derivation}.

We write done the explicit expressions of the reduced density operators:
\begin{eqnarray}
\label{eq:rhoAliceM}
&& \hat \varrho_A = \sum_{ik} \rho_{ik}\ket{\psi^A_i}\bra{\psi^A_k},
\quad \rho_{ik} = \sum_m c_{im}c^*_{km},
\\
\label{eq:rhoBobM}
&& \hat \varrho_B = \sum_{ik} \rho_{ik} \ket{\psi^B_i} \bra{\psi^B_k},
\quad \rho_{ik} = \sum_m c_{mi}c^*_{mk},
\\
\label{eq:rhoAZM}
&& \hat \varrho_{Z,A} = \sum_{i} \rho_{ii} \ket{\psi^A_i} \bra{\psi^A_i},
\quad \rho_{ii} = \sum_m c_{mi}c^*_{mi},
\\
\label{eq:rhoNZM}
&& \hat \varrho_{Z,B} = \sum_{ik} \rho_{ik} \ket{\psi^B_i} \bra{\psi^B_k},
\quad \rho_{ik} =\sum_{m} c_{mi}c^*_{mk}.
\end{eqnarray}

The characteristic equation for the matrix in Eq. (\ref{eq:cFromRho}) is
\begin{equation}
\label{eq:characteristicABMatrix}
\lambda^3(\lambda-\vert C_1 \vert^2 - \vert C_2 \vert^2 -\vert C_3 \vert
- \vert C_4 \vert)=0, 
\end{equation}
and the eigenvalues  $\lambda_{1,2} = 0, 1$.
For the projected matrix $\hat \varrho_{Z,AB}$,
\begin{equation}
\label{eq:characteristicZABMatrix}
\lambda^2 (\lambda - \vert C_1\vert^2 - \vert C_2 \vert^2)
(\lambda - \vert C_3\vert^2 - \vert C_4 \vert^2)=0, 
\end{equation}
we deduce:
$\lambda_{1,2} = 0$,
$\lambda_3 =\vert C_1\vert^2 + \vert C_2 \vert^2$, 
$\lambda_4 =\vert C_3\vert^2 + \vert C_4 \vert^2$.
The similar procedure we apply to the other post-measurement density matrices 
$\hat\varrho_{X, AB}$ and transformed coefficients $C^{(\Phi)}$.

For the reduced matrices in Eqs. (\ref{eq:rhoAliceM}), (\ref{eq:rhoBobM}) and
(\ref{eq:rhoNZM}) the solutions are given by:
\begin{equation}
\label{eq:lambdaReduced}
\lambda_{1,2} = \frac 1 2 \big[ 
1 \pm \sqrt{\frac 1 4 - \vert C_1C^*_4 - C^*_2C_3 \vert^2} \big] ,
\end{equation}
and for the density matrix Eq. (\ref{eq:rhoAZM}),
\begin{equation}
\label{eq:lambdaAZM}
\lambda_1 = \vert C_1 \vert^2 + \vert C_2 \vert^2, \,
\lambda_2 = \vert C_3 \vert^2 + \vert C_4 \vert^2.
\end{equation}
We introduce the function:
\begin{equation}
\label{eq:hdefinition}
h(x) = - x \log x - (1-x) \log (1-x).
\end{equation}
The pairs of distinct roots of the characteristic polynomuals of the entire 
and reduced density matrices satisfy the condition $\lambda_1 + \lambda_2 =1$.
It is easy to see that $h(x)=h(1-x)$, therefore $h(\lambda_1)=h(\lambda_2)$
and we can use either of them to calculate the entropy.

The entropies corresponding to the bipartite and reduced density operators read:
\begin{eqnarray}
\label{eq:SAB}
&&S(\hat \varrho_{AB}) = h(0) = 0, 
\\
\label{eq:SA}
&& S(\hat\varrho_A) = S(\hat \varrho_B)  = h(\lambda),
\\
\label{eq:SCondAB}
&& S(A|B) = - h(\lambda),
\end{eqnarray}
The entropy corresponding to the bipartite density operator 
is zero meaning that the system is in the pure state. According to Eq. (\ref{eq:characteristicABMatrix})
$\lambda = \frac 1 2 \big[
1 + \sqrt{\frac 1 4 - \vert C_1C^*_4 - C^*_2C_3 \vert^2} \big]$.
For post-measurement operators, after measuring the $z$ projection of the spin \textbf{A}, expressions of the entropies read:
\begin{eqnarray}
\label{eq:SZAB}
&& S(\hat \varrho_{Z,AB}) = h(\mu),
\\
\label{eq:SZB}
&& S(\hat \varrho_{Z,B}) = h(\lambda),
\\
\label{eq:SZcondB}
&& S(Z|B) =  h(\mu)-h(\lambda),
\end{eqnarray}
where
\begin{eqnarray}
&& \mu = \vert C_1\vert^2 + \vert C_2 \vert^2,
\label{eq:mu_from_C}
\\
&& \lambda = \frac 1 2 \big[
1 + \sqrt{\frac 1 4 - \vert C_1C^*_4 - C^*_2C_3 \vert^2} \big].
\label{eq:lambda_from_C}
\end{eqnarray}

After measuring $X$ component of the spin \textbf{A}, entropies of the post-measurement density operators read: 
\begin{eqnarray}
\label{eq:SXAB}
&& S(\hat \varrho_{X,AB}) = h(\xi), 
\\
\label{eq:SXB}
&& S(\hat \varrho_{X,B}) = h(\zeta),
\\
\label{eq:SXcondB}
&& S(X|B) =  h(\xi) - h(\zeta).
\end{eqnarray}
Here we introduced notations
\begin{eqnarray}
&&\xi=\vert C^{(\Phi)}_1 \vert ^2+\vert C^{(\Phi)}_2 \vert^2
\label{eq:xi_from_C}
\\
&&\zeta = \frac 1 2 \big[1 + 
\sqrt{\frac 1 4 - \vert C^{(\Phi)}_1C^{(\Phi)*}_4- C^{(\Phi)*}_2
C^{(\Phi)}_3 \vert^2}\big],
\label{eq:zeta_from_C}
\end{eqnarray}
The coefficients $C^{(\Phi)}_i$ are given by 
Eqs. (\ref{eq:cphi1})-(\ref{eq:cphi4}), and the eigenvalues $\xi$ and $\zeta$
are found through the original coefficient $C_i$. For more details we refer to the appendix (See Eqs. (\ref{eq:sumcphi12})-(\ref{eq:ProjectedMatrixPhi}).

Inserting all these expressions into Eq. (\ref{eq:BobsIgnorance})
we obtain the following form of the inequality:
\begin{equation}
\label{eq:BobsIgnorance2}
h(\xi) - h(\zeta) + h(\mu) 
\geq\log(1/c),
\end{equation}
or
\begin{eqnarray}
\label{eq:BobsIgnorance3}
 && -\xi \log\xi - (1-\xi) \log (1-\xi) +
\nonumber
\\
&& \quad
\zeta \log \zeta +  (1-\zeta) \log (1-\zeta)- 
\nonumber
\\
&& 
\quad \mu \log\mu -(1-\mu) \log (1-\mu) 
\geq \log (1/c).
\end{eqnarray}

\begin{figure}[h!]
\includegraphics[width=\columnwidth]{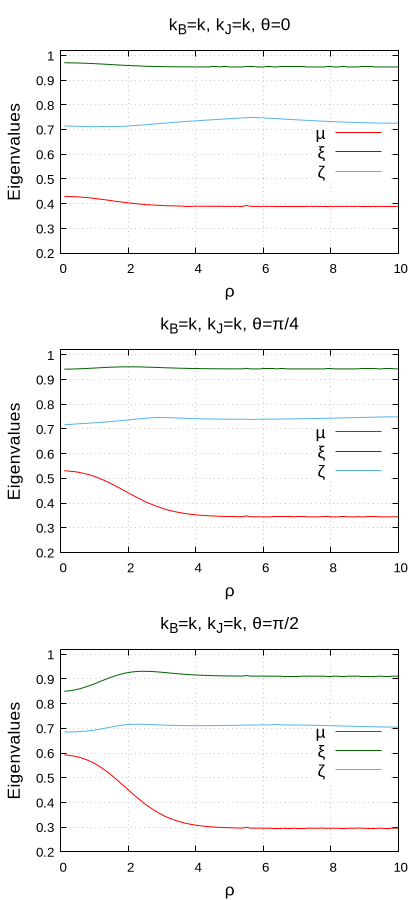}
\caption{ 
\label{fig:eigenvalues}
The eigenvalues $\mu$, $\zeta$ and $\xi$ of the post-measurement density matrices 
$\varrho_{Z,AB}$, $\varrho_{X,AB}$ and $\varrho_{X,B}$, respectively,
expressed by 
Eqs. (\ref{eq:mu_from_C}), (\ref{eq:zeta_from_C}) and (\ref{eq:xi_from_C}).
The eigenvalues are found for $k_B,k_J=k$ and three different values
of $\theta$.
}
\end{figure}

\begin{figure}[h!]
\includegraphics[width=\columnwidth]{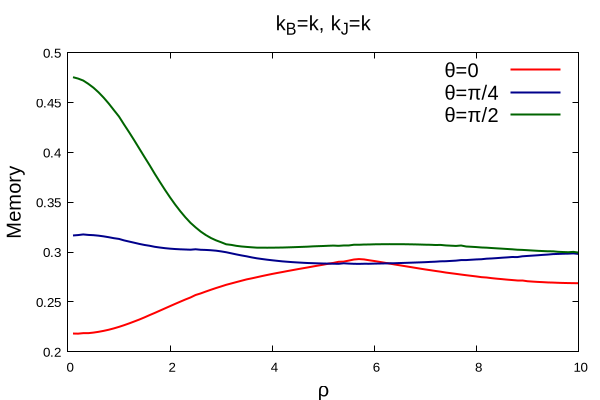}
\caption{ 
\label{fig:memory}
Quantum Memory calculated from 
Eqs.(\ref{eq:BobsIgnorance2}) and (\ref{eq:BobsIgnorance3})
for varying values of $k_B, k_J= k$ and three different values of $\theta$.
}
\end{figure}

We calculate the eigenvalues $\mu$, $\zeta$ and $\xi$ we calculate numerically for different values of $k$, $k_B$ and $k_J$.
Fig. \ref{fig:eigenvalues} represents dependence of $\mu$, $\zeta$ and $\xi$
on $\rho$ for $k=1$, $k_B=0.5$, $k_J=1$, $\theta=\pi/4$.
We quantify the quantum memory in terms of the difference between the left and the right hand sides of the equation Eq.(\ref{eq:BobsIgnorance}).
The memory calculated from Eqs.(\ref{eq:BobsIgnorance2}) and
(\ref{eq:BobsIgnorance3}) for three values $k_B=0.25$, $0.5$ and $1$, is
plotted in Fig. \ref{fig:memory}. 
As we see, the dependence of the quantum memory on the distance between the itinerant and localized electrons is distinct compared to the improved uncertainty relations for momentum and coordinate operators. While the momentum and coordinate operators' uncertainty increases with the distance between the electrons, the quantum memory does not decay, meaning that the uncertainty of spin measurements does not enhance at larger distances. 

\section{Conclusions}

It is well-known that the measurements done on the quantum systems are specific and may be invasive and destroy the state of interest. Beyond that, quantumness limits the accuracy of measurements due to the fundamental uncertainty relations. On the other hand, quantum correlations and memory might reduce quantum measurements' uncertainty. We studied two types of measurements done on the topological system: (a) measurements done on the spin operators and (b) measurements of the canonical pair of operators: momentum and coordinate. We quantified the spin operator's measurements through the entropic measures of uncertainty and exploited the concept of quantum memory. For the momentum and coordinate operators, we exploit the improved uncertainty relations. We showed that the dependence of the quantum memory on the distance between the itinerant and localized electrons is distinct compared to the improved uncertainty relations for momentum and coordinate operators. While the momentum and coordinate operators' uncertainty increases with the distance between the electrons, the quantum memory does not decay, meaning that the uncertainty of spin measurements does not enhance at elevated distances. Therefore, based on the discovered effect, we propose the indirect measurement scheme for the momentum and coordinate operators through the spin operator. Due to the factor of quantum memory, such indirect measurements in topological insulators lead to smaller uncertainties rather than direct measurements.


\appendix

\section{Coefficients of the wave function}
\label{sec:Coefficients}

\begin{widetext}
\begin{eqnarray}\label{notations for the coefficients}
&& A_{0}(\mathbf{r})=\int \rho'd\rho'd\varphi'\,K_0(-iE\vert\mathbf{r}-
\mathbf{r}'\vert/v)\vert\psi_D(\rho')\vert^2e^{ik\rho'\cos\varphi'},
\label{eq:A0}\\
&& A_{1}(\mathbf{r})=\int \rho'd\rho'd\varphi'\,\,K_1(-iE
\vert\mathbf{r}-\mathbf{r}'\vert/v)\vert\psi_D(\rho')\vert^2
e^{ik\rho'\cos\varphi'},\label{eq:A1}\\
&&A_{0B}(\mathbf{r})=\int \rho'd\rho'd\varphi'\,K_0(-i(E-2B)\vert\mathbf{r}-\mathbf{r}'\vert/v)\vert\psi_D(\rho')\vert^2e^{ik\rho'\cos\varphi'},\label{eq:A0B}\\
&&A_{1B}(\mathbf{r})=\int \rho'd\rho'd\varphi'\,\,K_1(-i(E-2B)\vert\mathbf{r}-\mathbf{r}'\vert/v)\vert\psi_D(\rho')\vert^2e^{ik\rho'\cos\varphi'}.\label{eq:A1B}
\end{eqnarray}
\end{widetext}
The explicit form of the matrix elements are:
\begin{widetext}
\begin{align}\label{The potential operators}
\hat{V}_{00}(\textbf{r}_A)&=\langle\psi_{D,0}(\textbf{r}_B)|\hat{V}|\psi_{D,0}(\textbf{r}_B)\rangle=\langle\psi_{D,0}(\textbf{r}_B)|J\hat\sigma_A\hat\sigma_B\delta\left(\textbf{r}_A-\textbf{r}_B\right)|\psi_{D,0}(\textbf{r}_B)\rangle=J|\psi_D(\textbf{r}_A)|^2\hat\sigma_A^z,\nonumber\\
\hat{V}_{01}(\textbf{r}_A)&=\langle\psi_{D,0}(\textbf{r}_B)|\hat{V}|\psi_{D,1}(\textbf{r}_B)\rangle=\langle\psi_{D,0}(\textbf{r}_B)|J\hat\sigma_A\hat\sigma_B\delta\left(\textbf{r}_A-\textbf{r}_B\right)|\psi_{D,1}(\textbf{r}_B)\rangle=J|\psi_D(\textbf{r}_A)|^2(\hat\sigma_A^x-i\hat\sigma_A^y),\nonumber\\
\hat{V}_{10}(\textbf{r}_A)&=\langle\psi_{D,1}(\textbf{r}_B)|\hat{V}|\psi_{D,0}(\textbf{r}_B)\rangle=\langle\psi_{D,1}(\textbf{r}_B)|J\hat\sigma_A\hat\sigma_B\delta\left(\textbf{r}_A-\textbf{r}_B\right)|\psi_{D,0}(\textbf{r}_B)\rangle=J|\psi_D(\textbf{r}_A)|^2(\hat\sigma_A^x+i\hat\sigma_A^y),\\
\hat{V}_{11}(\textbf{r}_A)&=\langle\psi_{D,1}(\textbf{r}_B)|\hat{V}|\psi_{D,1}(\textbf{r}_B)\rangle=\langle\psi_{D,1}(\textbf{r}_B)|J\hat\sigma_A\hat\sigma_B\delta\left(\textbf{r}_A-\textbf{r}_B\right)|\psi_{D,1}(\textbf{r}_B)\rangle=-J|\psi_D(\textbf{r}_A)|^2\hat\sigma_A^z.\nonumber
\end{align}
\end{widetext}

\vspace{20pt}

\section{Approximating integrals}

We utilize the following asymptotic for Bessel functions at $z\gg 1$:
\begin{equation}
\label{eq:Bessel_approx}
K_{l=0,1}(z)=\frac{\pi}{2z}e^{-z}\lbrace 1+\frac{4l^2-1}{8z}\rbrace,
\end{equation}
$z
=-ik_{(B)}|\textbf{r}-\textbf{r}'|=
-ik_{(B)}\sqrt{\rho^2+\rho'^2-2\rho\rho'\cos(\varphi-\varphi')}$,
$k_{(B)} =\frac {E-(2B)} v = k - (2B/v)$.
We apply these asymptotic expressions to Eq. (\ref{eq:A0}) and (\ref{eq:A1}):
\begin{eqnarray}
\nonumber
&&A_{l=0,1(B)}(\vec r ) \approx 
\int \rho'd\rho'  d\gamma \vert \psi_D(\rho')\vert^2
e^{ik\rho'\cos (\gamma+\varphi)} \times
\\
&& \qquad \qquad \qquad \frac \pi {2z}e^{-z} \left(1 + \frac {4l^2 -1} {8z}\right),
\label{eq:A0_approx}
\end{eqnarray}
where $\gamma = \varphi' - \varphi$ is the scattering angle.
We express $z$ as
$z = -ik_{(B)} a \sqrt{1 - b\cos \gamma}$, where
$a = \sqrt{\rho^2 + \rho'^2}$ and
$b =2 \rho \rho'/a^2$.
Assuming $b$ is a small parameter, we expand square root into power series and retain only leading order terms. Performing integration over the angle $\gamma$
we calculate the following integral:
\begin{eqnarray}
\nonumber
&&{I}_\gamma = -\int_0^{2\pi}d\gamma  \exp[ik\rho' \cos(\gamma + \varphi) +
ik_{(B)}a(1 - \frac b 2 \cos \gamma)] \times
\\
&& \qquad  
\left [ \frac 1 {ik_{(B)}a} \frac 1 { (1 - \frac b 2 \cos\gamma)}
+ \frac {4l^2-1} {8 k_{(B)}^2a^2 } \frac 1 {(1 -b \cos \gamma )}\right].
\label{eq:integral_by_gamma}
\end{eqnarray}
Both integrals in this expression can be transformed into contour integrals along
a circle of unit radius around the origin:
Assuming $z = e^{i\gamma}$, $d\gamma = (1/i)dz/z$, $\cos \gamma =(1/2)(z+ 1/z)$,
we deduce
\begin{equation}
\label{eq:OneOfTheIntegrals}
\int_{\vert z \vert=1} dz \frac{f(z)} {2z - cz^2 -c}.
\end{equation}
Here $c$ is either $b$ or $b/2$. 
The square polynomial in the denominator has two real roots:
\begin{equation}
\label{eq:roots}
z_{1,2} = \frac 1 c [1 \pm  \sqrt{1-c^2}].
\end{equation}
The only root with the  minus sign is located inside the unit circle 
at $z_0 \approx c/2 e$ and $\arg z_0 =0$ and both residuals are equal to
$-1$. 
The final form of Eq. (\ref{eq:integral_by_gamma}) reads
\begin{eqnarray}
\nonumber
&&{I}_\gamma =  \pi \frac {-8ik_{(B)}a +4l^2 -1}{4k_{(B)}^2a^2} \times
\\
&& \qquad \qquad \exp\left[ik\rho' \cos \varphi + ik_{(B)}a (1 -  b/2)
\right],
\label{eq:integral_by_gamma2}
\end{eqnarray}
and 
\begin{eqnarray}
\nonumber
&& A_{l=0,1(B)}(\vec r)  = \frac {\pi} { 4 l_B^2} \int \rho' d \rho' 
\frac {-8i k_{(B)} \sqrt{\rho^2 + \rho'^2} +4l^2 -1}
{ k_{(B)}^2 (\rho^2+\rho'^2)} \times
\\
&& \quad \exp \big[ \frac{ -\rho'^2 }{l_B^2} + i k\rho' \cos \varphi	
+ik_{(B)}\frac{{\rho^2 +\rho'^2}- {\rho \rho'}}
{\sqrt{\rho ^2 + \rho'^2}}\big].
\label{eq:A_after_varphiprime_integration}
\end{eqnarray}
Introducing notation $x = \rho'/\rho$, we rewrite the expression as follows
\begin{eqnarray}
\nonumber
&& A_{l=0,1(B)} (\vec r)= \frac \pi {4l_B^2} \int x dx  
\frac{-8 i k_{(B)}\rho \sqrt{1+x^2}+4l^2-1}{k_{(B)}^2 \rho^2(1+x^2)}\times
\\
\nonumber
&& \exp \left[-\left(\frac \rho {l_B}\right)^2x^2 +
ik\rho x \cos \varphi + ik_{(B)}\rho\frac{1-x+x^2}{\sqrt{1+x^2}} \right]\approx 
\\
\nonumber
&& \qquad \frac \pi {4l_B^2} \frac 1 {k_{(B)}^2 \rho^2} \int x dx 
\big[ \left(4ik_{(B)}\rho - 4l^2 +1\right)x^2 - 
\\
\nonumber
&& \qquad \qquad \qquad \left(8ik_{(B)} \rho -4l^2 +1 \right)\big] \times
\\
\nonumber
&& \qquad \exp\big[ - \left(\frac {\rho^2}{l_B^2} + \frac{3i} 2 \right) x^2
+i(k\cos \varphi - k_{(B)}) \rho x  +
\\
&& \qquad \qquad \qquad  \qquad \qquad  ik_{(B)}\rho\big].
\label{eq:A_after_varphiprime_integration2}
\end{eqnarray}
In the next step we analyze the term proportional to $x^3$. Considering that
$\rho^2/l_B^2 \gg 1 $ we perform the integration over the $x$ and obtain
\begin{eqnarray}
\nonumber
&&A_{l=0,1}  (\vec r) \approx \frac \pi {8 (k_{(B)}l_B)} 
\frac {8ik_{(B)}\rho - 4l^2 +1 } {(k_{(B)} \rho)^3} 
k_{(B)}^2{e^{ik_{(B)}\rho}} \times
\\
&&  \qquad   \left[1 + \sqrt{\pi} {i \kappa_{(B)} l_B} 
\left[ 1 + \erf({i\kappa_{(B)}l_B}) \right] e^{-\kappa^2 l_B^2}\right]
\label{eq:after_integration_by_x},
\end{eqnarray}
where $\kappa_{(B)} = ({k\cos \varphi - k_{B}})/ 2 $.
Note that the last line in Eq. (\ref{eq:after_integration_by_x})
 is a function of $\varphi$ and expand it in terms 
of the small parameter $ l_B\kappa_{(B)} \ll 1$:
\begin{eqnarray}
\nonumber
&& 1 + \sqrt{\pi} {i \kappa_{(B)} l_B} 
\left[ 1 + \erf({i\kappa_{(B)}l_B}) \right] e^{-\kappa^2 l_B^2}  \approx
\\
\nonumber
&& \qquad 1 + i \sqrt \pi \kappa_{(B)}l_B -2(\kappa_{(B)}l_B)^2 
= 1 - \frac {k_{(B)}^2l_B^2}2  -
\\
\nonumber
&& \qquad   \frac{i\sqrt \pi} 2 k_{(B)}l_B 
+\left(\frac{i\sqrt \pi} 2 + k_{(B)}l_B\right)kl_B \cos \varphi -
\\
&& \qquad \qquad  
\frac 1 2 k^2 l_B^2 \cos^2 \varphi.
\label{eq:varphi_approx}
\end{eqnarray}

\section{Expressions of integrals integrated over the angle}
\label{sec:expressions_for_integrals}

After integration over the angular variable $\varphi$, from Eq. (\ref{eq:Ivarphi}) we deduce:

\begin{eqnarray}
&&I^\varphi_{00} = \frac{19\pi}{16} k^4+\left(\frac 3 4 \pi -1 \right)\pi k^2 +
2\pi,
\label{eq:I00}
\\
\nonumber 
&& I^\varphi_{BB} = I^{\varphi}_{00} + 8\pi k_B^4 - 16\pi k k_B^3 +
(14 k^2+2\pi -
\\
&& \qquad 8)\pi k_{B}^2 + \left[(8 -2\pi)k - 6 k^3 \right]\pi k_{B},
\label{eq:IBB}
\\
\nonumber
&& I^\varphi_{0B} = I^\varphi_{00} - (4-k^2)\pi k_B^2 -
\left[3k^2 + (\pi-4) \right]\pi k  k_B-
\\
&& \qquad 2 i \pi\sqrt{\pi} \left[  k k_B^2 - \left(\frac 5 4 k^2 - 1 
\right)k_B\right],
\label{eq:I0B}
\\
&&  I^\varphi_{B0} =  (I^\varphi_{0B} )^*. 
\end{eqnarray}

The difference between Eqs. (\ref{eq:I00}) and  (\ref{eq:IBB}) vanishes
when $k_B=0, k$, and the expressions become symmetric with respect to $k_B=0.5k$.

\begin{figure}[h!]
\includegraphics[width=\columnwidth]{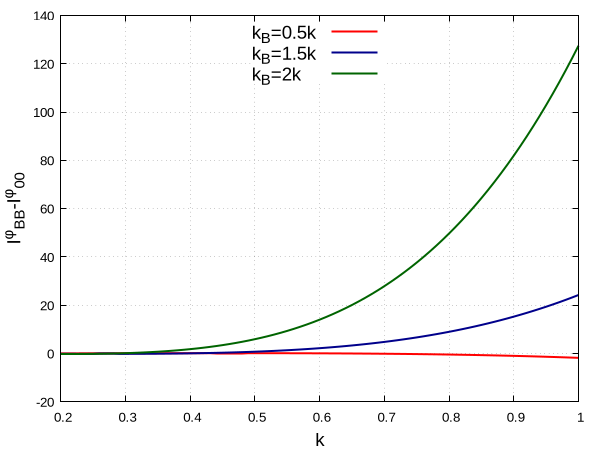}
\caption{
\label{fig:IBB}
Difference $I^\varphi_{BB} - I^\varphi_{00}$ as a function of $k$ for different
proportionality coefficients of $k_B$ relative to $k$.
}
\end{figure}

\section{Normalization of the wave function}
\label{sec:normalization}

We calculate integrals for the products of the functions $f_{\lambda b}$ used in the normalization of the wave function:
\begin{eqnarray}
\nonumber 
&& I^N_{\lambda_1 \lambda_2 b_1 b_2} = \int_1^{\infty} \rho d\rho
f^*_{\lambda_1b_1}(\rho)f_{\lambda_2b_2}(\rho) = \left[\frac \pi {k_{b_1}k_{b_2}}\right]^2\times
\\
&&  
\sum_{n=2}^4 G_ni^{n-1} (-i\Delta k_{(b)})^n\Gamma(-n, -i\Delta k_{(b)}),
\label{eq:IN}
\end{eqnarray}
where
\begin{eqnarray}
&& G_2 = k_{b_1}k_{b_2},
\label{eq:G2}
\\
&& G_3 = \frac 1 8 \left(\lambda_1 k_{b_2} -\lambda_2 k_{b_1}\right),
\label{eq:G3}
\\
&& G_4 = \frac{\lambda_1\lambda_2}{64}.
\label{eq:G4}
\end{eqnarray}

Expressions for all possible combinations of the  parameters read:
\begin{eqnarray}
&& I^N_{00,00} =  \frac {\pi^2}{2k^4}\left(k^2+\frac 1{128} \right),
\label{eq:I_0000}
\\
&& I^N_{11, 00} =  \frac {\pi^2}{2k^4}\left(k^2 + \frac 9 {128} \right),
\label{eq:I_1100}
\\
&&
I^N_{00,BB} =  \frac {\pi^2}{(k-2k_B)^4}\left((k-2k_B)^2+\frac 1{128} \right),
\label{eq:I_00BB}
\\
&& I^N_{11, BB} =\frac {\pi^2}{(k-2k_B)^4}\left((k-2k_B)^2 +\frac 9{128}\right),
\label{eq:I_1100}
\\
&& I^N_{10,00} =  \frac {\pi^2}{2k^4} \left(k^2 - \frac 3 {128} + \frac 1 3 ik
\right),
\label{eq:I_1000}
\\
&& I^N_{01,00} =   (I^N_{10,00})^*,
\label{eq:I_0100}
\\
&& I^N_{10,BB} =  \frac {\pi^2}{2(k-2k_B)^4} \big((k-2k_B)^2 - 
\frac 3 {128} + 
\nonumber 
\\
&& \qquad \qquad \frac 1 3 i(k-2k_B)
\big),
\label{eq:I_10BB}
\\
&& I^N_{01,BB} =  \frac {\pi^2}{2(k-2k_B)^4} \big((k-2k_B)^2 - 
\frac 3 {128} - 
\nonumber 
\\
&& \qquad \qquad \frac 1 3 i(k-2k_B)
\big),
\label{eq:I_01BB}
\end{eqnarray}

\begin{figure}[h!]
\includegraphics[width=\columnwidth]{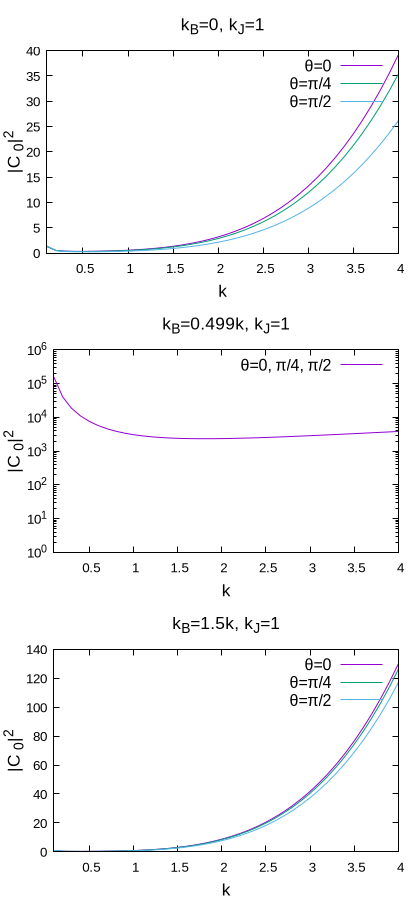}
\caption{\label{fig:C0} The dependence of $\vert C_0\vert^2$ on $k$ given in
Eq. (\ref{eq:C0_square}) and plotted for several $k_B$ and $\theta$ for $k_J=k$.
The function has a singularity when $k_B$ approaches $0.5k$.
}
\end{figure}

Squares of modules of the coefficients:
\begin{eqnarray}
\nonumber
&& \int \vert C_1\vert^2 =A (k_Jk)^2I^\varphi_{00} \big(
I^N_{00,00}+I^N_{11,00}+I^N_{01,00}e^{i\theta}+
\\
\nonumber 
&& \qquad I^N_{10,00}e^{-i\theta} \big)=A I^\varphi_{00}\frac{(\pi k_J)^2}{k^2}
\big[k^2 (1+\cos \theta) + 
\\
&& \qquad \frac 1 {128} (10-3\cos \theta) + \frac 1 3 \sin \theta \big],
\label{eq:C1_square}
\end{eqnarray}
where $A= \vert\psi_D(\rho')\vert^2/(8\pi^2)$ 
includes the rest of the normalization terms in front of the parenthesis.
Calculation of $\vert C_2\vert^2$ and $\vert C_4\vert^2$ is rather
straightforward. The expression for $\vert C_3\vert^2$ is different from Eq. (\ref{eq:C1_square}) 
only by sign in front of the $\sin\theta$ term.
Finally the normalization coefficient reads
\begin{eqnarray}
\nonumber
&& 
\vert C_0 \vert^2 =\int \sum_{i=1}^4  \vert C_i\vert^2=2A(\pi k_J)^2\times
\\
\nonumber
&& \qquad \big[\frac{I^\varphi_{00}}{k^2} \left(k^2(1+\cos\theta) +\frac 1{128}
(10-3\cos \theta)\right) + 
\\
&& \qquad \qquad \frac{I^\varphi_{BB}}{(k-2k_B)^2}
\left((k-2k_B)^2 + \frac 5 {64} \right) \big].
\label{eq:C0_square}
\end{eqnarray}
One can see, that the norm diverges at $k=0$ and  $k_B=0.5k$.
Dependence of $\vert C_0\vert^2$ on $k$ is plotted in Fig. \ref{fig:C0}.

\section{Coefficients in the new basis}
\label{sec:app_derivation}

Coefficients of the alternative representation of the wave function in the
basis $\{\phi_1, \phi_2\}$:

\begin{eqnarray}
&&\ket{\Phi}_{AB} =  C_1^{(\Phi)} \ket{\phi^A_1\phi^A_1} +
C_2^{(\Phi)} \ket{\phi^A_1\phi^A_2} +
\nonumber
\\
\label{eq:phiFunc}
&& \qquad \qquad
C_3^{(\Phi)} \ket{\phi^A_2\phi^A_1} +
C_4^{(\Phi)} \ket{\phi^A_2\phi^A_2},
\end{eqnarray}
where
\begin{eqnarray}
\label{eq:cphi1}
&&C_1^{(\Phi)} = \frac 1 2 \left(C_1 + C_2 + C_3 + C_4 \right),
\\
\label{eq:cphi2}
&&C_2^{(\Phi)} = \frac 1 2 \left(C_1 - C_2 + C_3 - C_4\right),
\\
\label{eq:cphi3}
&& C_3^{(\Phi)} = \frac 1 2 \left(C_1 + C_2 - C_3  - C_4 \right),
\\
\label{eq:cphi4}
&& C_4^{(\Phi)}= \frac 1 2 \left(C_1 - C_2 - C_3 + C_4 \right).
\end{eqnarray}

One can show that 

\begin{eqnarray}
&& \vert C_1^{(\Phi)}\vert^2 + \vert C_2^{(\Phi)}\vert^2 =
\frac 1 2 \big[1 + (C_1C^*_3 + C^*_1C_3) + 
\nonumber
\\
\label{eq:sumcphi12}
&& \qquad \qquad \qquad
(C_2C^*_4 +C^*_2C_4)\big],
\\
&& \vert C_3^{(\Phi)}\vert^2 + \vert C_4^{(\Phi)}\vert^2 =
\frac 1 2 \big[1 - (C_1C^*_3 + C^*_1C_3) -\nonumber
\\
\label{eq:sumcphi34}
&& \qquad \qquad \qquad 
(C_2C^*_4 +C^*_2C_4)\big],
\\
&& C_1^{(\Phi)}C_4^{(\Phi)*} -  C_2^{(\Phi)*}C_3^{(\Phi)} =
\frac 1 2 \big[1-(C_1+C_2)(C^*_3-C^*_4) +
\nonumber
\\
\label{eq:ProjectedMatrixPhi}
&& \qquad \qquad \qquad 
(C^*_1-C^*_2)(C_3+C_4)\big].
\end{eqnarray}

\section*{Acknowledgment}

\bibliography{TEXT_KK_VD.LC}

\begin{thebibliography}{24}%
\makeatletter
\providecommand \@ifxundefined [1]{%
 \@ifx{#1\undefined}
}%
\providecommand \@ifnum [1]{%
 \ifnum #1\expandafter \@firstoftwo
 \else \expandafter \@secondoftwo
 \fi
}%
\providecommand \@ifx [1]{%
 \ifx #1\expandafter \@firstoftwo
 \else \expandafter \@secondoftwo
 \fi
}%
\providecommand \natexlab [1]{#1}%
\providecommand \enquote  [1]{``#1''}%
\providecommand \bibnamefont  [1]{#1}%
\providecommand \bibfnamefont [1]{#1}%
\providecommand \citenamefont [1]{#1}%
\providecommand \href@noop [0]{\@secondoftwo}%
\providecommand \href [0]{\begingroup \@sanitize@url \@href}%
\providecommand \@href[1]{\@@startlink{#1}\@@href}%
\providecommand \@@href[1]{\endgroup#1\@@endlink}%
\providecommand \@sanitize@url [0]{\catcode `\\12\catcode `\$12\catcode
  `\&12\catcode `\#12\catcode `\^12\catcode `\_12\catcode `\%12\relax}%
\providecommand \@@startlink[1]{}%
\providecommand \@@endlink[0]{}%
\providecommand \url  [0]{\begingroup\@sanitize@url \@url }%
\providecommand \@url [1]{\endgroup\@href {#1}{\urlprefix }}%
\providecommand \urlprefix  [0]{URL }%
\providecommand \Eprint [0]{\href }%
\providecommand \doibase [0]{https://doi.org/}%
\providecommand \selectlanguage [0]{\@gobble}%
\providecommand \bibinfo  [0]{\@secondoftwo}%
\providecommand \bibfield  [0]{\@secondoftwo}%
\providecommand \translation [1]{[#1]}%
\providecommand \BibitemOpen [0]{}%
\providecommand \bibitemStop [0]{}%
\providecommand \bibitemNoStop [0]{.\EOS\space}%
\providecommand \EOS [0]{\spacefactor3000\relax}%
\providecommand \BibitemShut  [1]{\csname bibitem#1\endcsname}%
\let\auto@bib@innerbib\@empty
\bibitem [{\citenamefont {Messiah}(2014)}]{messiah2014quantum}%
  \BibitemOpen
  \bibfield  {author} {\bibinfo {author} {\bibfnamefont {A.}~\bibnamefont
  {Messiah}},\ }\href@noop {} {\emph {\bibinfo {title} {Quantum mechanics}}}\
  (\bibinfo  {publisher} {Courier Corporation},\ \bibinfo {year}
  {2014})\BibitemShut {NoStop}%
\bibitem [{\citenamefont {Coles}\ \emph {et~al.}(2017)\citenamefont {Coles},
  \citenamefont {Berta}, \citenamefont {Tomamichel},\ and\ \citenamefont
  {Wehner}}]{RevModPhys.89.015002}%
  \BibitemOpen
  \bibfield  {author} {\bibinfo {author} {\bibfnamefont {P.~J.}\ \bibnamefont
  {Coles}}, \bibinfo {author} {\bibfnamefont {M.}~\bibnamefont {Berta}},
  \bibinfo {author} {\bibfnamefont {M.}~\bibnamefont {Tomamichel}},\ and\
  \bibinfo {author} {\bibfnamefont {S.}~\bibnamefont {Wehner}},\ }\bibfield
  {title} {\bibinfo {title} {Entropic uncertainty relations and their
  applications},\ }\href {https://doi.org/10.1103/RevModPhys.89.015002}
  {\bibfield  {journal} {\bibinfo  {journal} {Rev. Mod. Phys.}\ }\textbf
  {\bibinfo {volume} {89}},\ \bibinfo {pages} {015002} (\bibinfo {year}
  {2017})}\BibitemShut {NoStop}%
\bibitem [{\citenamefont {Maccone}\ and\ \citenamefont
  {Pati}(2014)}]{PhysRevLett.113.260401}%
  \BibitemOpen
  \bibfield  {author} {\bibinfo {author} {\bibfnamefont {L.}~\bibnamefont
  {Maccone}}\ and\ \bibinfo {author} {\bibfnamefont {A.~K.}\ \bibnamefont
  {Pati}},\ }\bibfield  {title} {\bibinfo {title} {Stronger uncertainty
  relations for all incompatible observables},\ }\href
  {https://doi.org/10.1103/PhysRevLett.113.260401} {\bibfield  {journal}
  {\bibinfo  {journal} {Phys. Rev. Lett.}\ }\textbf {\bibinfo {volume} {113}},\
  \bibinfo {pages} {260401} (\bibinfo {year} {2014})}\BibitemShut {NoStop}%
\bibitem [{\citenamefont {Berta}\ \emph {et~al.}(2010)\citenamefont {Berta},
  \citenamefont {Christandl}, \citenamefont {Colbeck}, \citenamefont {Renes},\
  and\ \citenamefont {Renner}}]{berta2010uncertainty}%
  \BibitemOpen
  \bibfield  {author} {\bibinfo {author} {\bibfnamefont {M.}~\bibnamefont
  {Berta}}, \bibinfo {author} {\bibfnamefont {M.}~\bibnamefont {Christandl}},
  \bibinfo {author} {\bibfnamefont {R.}~\bibnamefont {Colbeck}}, \bibinfo
  {author} {\bibfnamefont {J.~M.}\ \bibnamefont {Renes}},\ and\ \bibinfo
  {author} {\bibfnamefont {R.}~\bibnamefont {Renner}},\ }\bibfield  {title}
  {\bibinfo {title} {The uncertainty principle in the presence of quantum
  memory},\ }\href@noop {} {\bibfield  {journal} {\bibinfo  {journal} {Nature
  Physics}\ }\textbf {\bibinfo {volume} {6}},\ \bibinfo {pages} {659} (\bibinfo
  {year} {2010})}\BibitemShut {NoStop}%
\bibitem [{\citenamefont {Wang}\ \emph {et~al.}(2019)\citenamefont {Wang},
  \citenamefont {Ming}, \citenamefont {Hu},\ and\ \citenamefont
  {Ye}}]{wang2019quantum}%
  \BibitemOpen
  \bibfield  {author} {\bibinfo {author} {\bibfnamefont {D.}~\bibnamefont
  {Wang}}, \bibinfo {author} {\bibfnamefont {F.}~\bibnamefont {Ming}}, \bibinfo
  {author} {\bibfnamefont {M.-L.}\ \bibnamefont {Hu}},\ and\ \bibinfo {author}
  {\bibfnamefont {L.}~\bibnamefont {Ye}},\ }\bibfield  {title} {\bibinfo
  {title} {Quantum-memory-assisted entropic uncertainty relations},\
  }\href@noop {} {\bibfield  {journal} {\bibinfo  {journal} {Annalen der
  Physik}\ }\textbf {\bibinfo {volume} {531}},\ \bibinfo {pages} {1900124}
  (\bibinfo {year} {2019})}\BibitemShut {NoStop}%
\bibitem [{\citenamefont {Ming}\ \emph {et~al.}(2020)\citenamefont {Ming},
  \citenamefont {Wang}, \citenamefont {Fan}, \citenamefont {Shi}, \citenamefont
  {Ye},\ and\ \citenamefont {Chen}}]{ming2020improved}%
  \BibitemOpen
  \bibfield  {author} {\bibinfo {author} {\bibfnamefont {F.}~\bibnamefont
  {Ming}}, \bibinfo {author} {\bibfnamefont {D.}~\bibnamefont {Wang}}, \bibinfo
  {author} {\bibfnamefont {X.-G.}\ \bibnamefont {Fan}}, \bibinfo {author}
  {\bibfnamefont {W.-N.}\ \bibnamefont {Shi}}, \bibinfo {author} {\bibfnamefont
  {L.}~\bibnamefont {Ye}},\ and\ \bibinfo {author} {\bibfnamefont {J.-L.}\
  \bibnamefont {Chen}},\ }\bibfield  {title} {\bibinfo {title} {Improved
  tripartite uncertainty relation with quantum memory},\ }\href@noop {}
  {\bibfield  {journal} {\bibinfo  {journal} {Physical Review A}\ }\textbf
  {\bibinfo {volume} {102}},\ \bibinfo {pages} {012206} (\bibinfo {year}
  {2020})}\BibitemShut {NoStop}%
\bibitem [{\citenamefont {Dolatkhah}\ \emph {et~al.}(2020)\citenamefont
  {Dolatkhah}, \citenamefont {Haseli}, \citenamefont {Salimi},\ and\
  \citenamefont {Khorashad}}]{dolatkhah2020tightening}%
  \BibitemOpen
  \bibfield  {author} {\bibinfo {author} {\bibfnamefont {H.}~\bibnamefont
  {Dolatkhah}}, \bibinfo {author} {\bibfnamefont {S.}~\bibnamefont {Haseli}},
  \bibinfo {author} {\bibfnamefont {S.}~\bibnamefont {Salimi}},\ and\ \bibinfo
  {author} {\bibfnamefont {A.}~\bibnamefont {Khorashad}},\ }\bibfield  {title}
  {\bibinfo {title} {Tightening the tripartite quantum-memory-assisted entropic
  uncertainty relation},\ }\href@noop {} {\bibfield  {journal} {\bibinfo
  {journal} {Physical Review A}\ }\textbf {\bibinfo {volume} {102}},\ \bibinfo
  {pages} {052227} (\bibinfo {year} {2020})}\BibitemShut {NoStop}%
\bibitem [{\citenamefont {Bergh}\ and\ \citenamefont
  {G{\"a}rttner}(2021)}]{bergh2021entanglement}%
  \BibitemOpen
  \bibfield  {author} {\bibinfo {author} {\bibfnamefont {B.}~\bibnamefont
  {Bergh}}\ and\ \bibinfo {author} {\bibfnamefont {M.}~\bibnamefont
  {G{\"a}rttner}},\ }\bibfield  {title} {\bibinfo {title} {Entanglement
  detection in quantum many-body systems using entropic uncertainty
  relations},\ }\href@noop {} {\bibfield  {journal} {\bibinfo  {journal}
  {Physical Review A}\ }\textbf {\bibinfo {volume} {103}},\ \bibinfo {pages}
  {052412} (\bibinfo {year} {2021})}\BibitemShut {NoStop}%
\bibitem [{\citenamefont {Xie}\ \emph {et~al.}(2021)\citenamefont {Xie},
  \citenamefont {Ming}, \citenamefont {Wang}, \citenamefont {Ye},\ and\
  \citenamefont {Chen}}]{PhysRevA.104.062204}%
  \BibitemOpen
  \bibfield  {author} {\bibinfo {author} {\bibfnamefont {B.-F.}\ \bibnamefont
  {Xie}}, \bibinfo {author} {\bibfnamefont {F.}~\bibnamefont {Ming}}, \bibinfo
  {author} {\bibfnamefont {D.}~\bibnamefont {Wang}}, \bibinfo {author}
  {\bibfnamefont {L.}~\bibnamefont {Ye}},\ and\ \bibinfo {author}
  {\bibfnamefont {J.-L.}\ \bibnamefont {Chen}},\ }\bibfield  {title} {\bibinfo
  {title} {Optimized entropic uncertainty relations for multiple
  measurements},\ }\href {https://doi.org/10.1103/PhysRevA.104.062204}
  {\bibfield  {journal} {\bibinfo  {journal} {Phys. Rev. A}\ }\textbf {\bibinfo
  {volume} {104}},\ \bibinfo {pages} {062204} (\bibinfo {year}
  {2021})}\BibitemShut {NoStop}%
\bibitem [{\citenamefont {Chotorlishvili}\ \emph {et~al.}(2019)\citenamefont
  {Chotorlishvili}, \citenamefont {Gudyma}, \citenamefont {W{\"a}tzel},
  \citenamefont {Ernst},\ and\ \citenamefont
  {Berakdar}}]{chotorlishvili2019spin}%
  \BibitemOpen
  \bibfield  {author} {\bibinfo {author} {\bibfnamefont {L.}~\bibnamefont
  {Chotorlishvili}}, \bibinfo {author} {\bibfnamefont {A.}~\bibnamefont
  {Gudyma}}, \bibinfo {author} {\bibfnamefont {J.}~\bibnamefont {W{\"a}tzel}},
  \bibinfo {author} {\bibfnamefont {A.}~\bibnamefont {Ernst}},\ and\ \bibinfo
  {author} {\bibfnamefont {J.}~\bibnamefont {Berakdar}},\ }\bibfield  {title}
  {\bibinfo {title} {Spin-orbit-coupled quantum memory of a double quantum
  dot},\ }\href@noop {} {\bibfield  {journal} {\bibinfo  {journal} {Physical
  Review B}\ }\textbf {\bibinfo {volume} {100}},\ \bibinfo {pages} {174413}
  (\bibinfo {year} {2019})}\BibitemShut {NoStop}%
\bibitem [{\citenamefont {Song}\ \emph {et~al.}(2022)\citenamefont {Song},
  \citenamefont {Li}, \citenamefont {Song}, \citenamefont {Ye},\ and\
  \citenamefont {Wang}}]{song2022environment}%
  \BibitemOpen
  \bibfield  {author} {\bibinfo {author} {\bibfnamefont {M.-L.}\ \bibnamefont
  {Song}}, \bibinfo {author} {\bibfnamefont {L.-J.}\ \bibnamefont {Li}},
  \bibinfo {author} {\bibfnamefont {X.-K.}\ \bibnamefont {Song}}, \bibinfo
  {author} {\bibfnamefont {L.}~\bibnamefont {Ye}},\ and\ \bibinfo {author}
  {\bibfnamefont {D.}~\bibnamefont {Wang}},\ }\bibfield  {title} {\bibinfo
  {title} {Environment-mediated entropic uncertainty in charging quantum
  batteries},\ }\href@noop {} {\bibfield  {journal} {\bibinfo  {journal}
  {Physical Review E}\ }\textbf {\bibinfo {volume} {106}},\ \bibinfo {pages}
  {054107} (\bibinfo {year} {2022})}\BibitemShut {NoStop}%
\bibitem [{\citenamefont {Zhu}(2021)}]{zhu2021zero}%
  \BibitemOpen
  \bibfield  {author} {\bibinfo {author} {\bibfnamefont {H.}~\bibnamefont
  {Zhu}},\ }\bibfield  {title} {\bibinfo {title} {Zero uncertainty states in
  the presence of quantum memory},\ }\href@noop {} {\bibfield  {journal}
  {\bibinfo  {journal} {npj Quantum Information}\ }\textbf {\bibinfo {volume}
  {7}},\ \bibinfo {pages} {47} (\bibinfo {year} {2021})}\BibitemShut {NoStop}%
\bibitem [{\citenamefont {Kurashvili}\ \emph {et~al.}(2022)\citenamefont
  {Kurashvili}, \citenamefont {Chotorlishvili}, \citenamefont {Kouzakov},\ and\
  \citenamefont {Studenikin}}]{kurashvili2022quantum}%
  \BibitemOpen
  \bibfield  {author} {\bibinfo {author} {\bibfnamefont {P.}~\bibnamefont
  {Kurashvili}}, \bibinfo {author} {\bibfnamefont {L.}~\bibnamefont
  {Chotorlishvili}}, \bibinfo {author} {\bibfnamefont {K.}~\bibnamefont
  {Kouzakov}},\ and\ \bibinfo {author} {\bibfnamefont {A.}~\bibnamefont
  {Studenikin}},\ }\bibfield  {title} {\bibinfo {title} {Quantum spin-flavour
  memory of ultrahigh-energy neutrino},\ }\href@noop {} {\bibfield  {journal}
  {\bibinfo  {journal} {The European Physical Journal Plus}\ }\textbf {\bibinfo
  {volume} {137}},\ \bibinfo {pages} {234} (\bibinfo {year}
  {2022})}\BibitemShut {NoStop}%
\bibitem [{\citenamefont {Kurashvili}\ \emph {et~al.}(2021)\citenamefont
  {Kurashvili}, \citenamefont {Chotorlishvili}, \citenamefont {Kouzakov},
  \citenamefont {Tevzadze},\ and\ \citenamefont
  {Studenikin}}]{PhysRevD.103.036011}%
  \BibitemOpen
  \bibfield  {author} {\bibinfo {author} {\bibfnamefont {P.}~\bibnamefont
  {Kurashvili}}, \bibinfo {author} {\bibfnamefont {L.}~\bibnamefont
  {Chotorlishvili}}, \bibinfo {author} {\bibfnamefont {K.~A.}\ \bibnamefont
  {Kouzakov}}, \bibinfo {author} {\bibfnamefont {A.~G.}\ \bibnamefont
  {Tevzadze}},\ and\ \bibinfo {author} {\bibfnamefont {A.~I.}\ \bibnamefont
  {Studenikin}},\ }\bibfield  {title} {\bibinfo {title} {Quantum witness and
  invasiveness of cosmic neutrino measurements},\ }\href
  {https://doi.org/10.1103/PhysRevD.103.036011} {\bibfield  {journal} {\bibinfo
   {journal} {Phys. Rev. D}\ }\textbf {\bibinfo {volume} {103}},\ \bibinfo
  {pages} {036011} (\bibinfo {year} {2021})}\BibitemShut {NoStop}%
\bibitem [{\citenamefont {Yazyev}\ \emph {et~al.}(2010)\citenamefont {Yazyev},
  \citenamefont {Moore},\ and\ \citenamefont {Louie}}]{PhysRevLett.105.266806}%
  \BibitemOpen
  \bibfield  {author} {\bibinfo {author} {\bibfnamefont {O.~V.}\ \bibnamefont
  {Yazyev}}, \bibinfo {author} {\bibfnamefont {J.~E.}\ \bibnamefont {Moore}},\
  and\ \bibinfo {author} {\bibfnamefont {S.~G.}\ \bibnamefont {Louie}},\
  }\bibfield  {title} {\bibinfo {title} {Spin polarization and transport of
  surface states in the topological insulators
  ${\mathrm{bi}}_{2}{\mathrm{se}}_{3}$ and ${\mathrm{bi}}_{2}{\mathrm{te}}_{3}$
  from first principles},\ }\href
  {https://doi.org/10.1103/PhysRevLett.105.266806} {\bibfield  {journal}
  {\bibinfo  {journal} {Phys. Rev. Lett.}\ }\textbf {\bibinfo {volume} {105}},\
  \bibinfo {pages} {266806} (\bibinfo {year} {2010})}\BibitemShut {NoStop}%
\bibitem [{\citenamefont {Seifert}\ \emph {et~al.}(2021)\citenamefont
  {Seifert}, \citenamefont {Kovarik}, \citenamefont {Gambardella},\ and\
  \citenamefont {Stepanow}}]{PhysRevResearch.3.043185}%
  \BibitemOpen
  \bibfield  {author} {\bibinfo {author} {\bibfnamefont {T.~S.}\ \bibnamefont
  {Seifert}}, \bibinfo {author} {\bibfnamefont {S.}~\bibnamefont {Kovarik}},
  \bibinfo {author} {\bibfnamefont {P.}~\bibnamefont {Gambardella}},\ and\
  \bibinfo {author} {\bibfnamefont {S.}~\bibnamefont {Stepanow}},\ }\bibfield
  {title} {\bibinfo {title} {Accurate measurement of atomic magnetic moments by
  minimizing the tip magnetic field in stm-based electron paramagnetic
  resonance},\ }\href {https://doi.org/10.1103/PhysRevResearch.3.043185}
  {\bibfield  {journal} {\bibinfo  {journal} {Phys. Rev. Research}\ }\textbf
  {\bibinfo {volume} {3}},\ \bibinfo {pages} {043185} (\bibinfo {year}
  {2021})}\BibitemShut {NoStop}%
\bibitem [{\citenamefont {Willke}\ \emph {et~al.}(2019)\citenamefont {Willke},
  \citenamefont {Singha}, \citenamefont {Zhang}, \citenamefont {Esat},
  \citenamefont {Lutz}, \citenamefont {Heinrich},\ and\ \citenamefont
  {Choi}}]{willke2019tuning}%
  \BibitemOpen
  \bibfield  {author} {\bibinfo {author} {\bibfnamefont {P.}~\bibnamefont
  {Willke}}, \bibinfo {author} {\bibfnamefont {A.}~\bibnamefont {Singha}},
  \bibinfo {author} {\bibfnamefont {X.}~\bibnamefont {Zhang}}, \bibinfo
  {author} {\bibfnamefont {T.}~\bibnamefont {Esat}}, \bibinfo {author}
  {\bibfnamefont {C.~P.}\ \bibnamefont {Lutz}}, \bibinfo {author}
  {\bibfnamefont {A.~J.}\ \bibnamefont {Heinrich}},\ and\ \bibinfo {author}
  {\bibfnamefont {T.}~\bibnamefont {Choi}},\ }\bibfield  {title} {\bibinfo
  {title} {Tuning single-atom electron spin resonance in a vector magnetic
  field},\ }\href@noop {} {\bibfield  {journal} {\bibinfo  {journal} {Nano
  Letters}\ }\textbf {\bibinfo {volume} {19}},\ \bibinfo {pages} {8201}
  (\bibinfo {year} {2019})}\BibitemShut {NoStop}%
\bibitem [{\citenamefont {Baruffa}\ \emph {et~al.}(2010)\citenamefont
  {Baruffa}, \citenamefont {Stano},\ and\ \citenamefont
  {Fabian}}]{PhysRevB.82.045311}%
  \BibitemOpen
  \bibfield  {author} {\bibinfo {author} {\bibfnamefont {F.}~\bibnamefont
  {Baruffa}}, \bibinfo {author} {\bibfnamefont {P.}~\bibnamefont {Stano}},\
  and\ \bibinfo {author} {\bibfnamefont {J.}~\bibnamefont {Fabian}},\
  }\bibfield  {title} {\bibinfo {title} {Spin-orbit coupling and anisotropic
  exchange in two-electron double quantum dots},\ }\href
  {https://doi.org/10.1103/PhysRevB.82.045311} {\bibfield  {journal} {\bibinfo
  {journal} {Phys. Rev. B}\ }\textbf {\bibinfo {volume} {82}},\ \bibinfo
  {pages} {045311} (\bibinfo {year} {2010})}\BibitemShut {NoStop}%
\bibitem [{\citenamefont {Cordourier-Maruri}\ \emph {et~al.}(2014)\citenamefont
  {Cordourier-Maruri}, \citenamefont {Omar}, \citenamefont {de~Coss},\ and\
  \citenamefont {Bose}}]{PhysRevB.89.075426}%
  \BibitemOpen
  \bibfield  {author} {\bibinfo {author} {\bibfnamefont {G.}~\bibnamefont
  {Cordourier-Maruri}}, \bibinfo {author} {\bibfnamefont {Y.}~\bibnamefont
  {Omar}}, \bibinfo {author} {\bibfnamefont {R.}~\bibnamefont {de~Coss}},\ and\
  \bibinfo {author} {\bibfnamefont {S.}~\bibnamefont {Bose}},\ }\bibfield
  {title} {\bibinfo {title} {Graphene-enabled low-control quantum gates between
  static and mobile spins},\ }\href
  {https://doi.org/10.1103/PhysRevB.89.075426} {\bibfield  {journal} {\bibinfo
  {journal} {Phys. Rev. B}\ }\textbf {\bibinfo {volume} {89}},\ \bibinfo
  {pages} {075426} (\bibinfo {year} {2014})}\BibitemShut {NoStop}%
\bibitem [{\citenamefont {Wolski}\ \emph {et~al.}(2022)\citenamefont {Wolski},
  \citenamefont {Inglot}, \citenamefont {Jasiukiewicz}, \citenamefont
  {Kouzakov}, \citenamefont {Mas\l{}owski}, \citenamefont
  {Szczepa\ifmmode~\acute{n}\else \'{n}\fi{}ski}, \citenamefont
  {Stagraczy\ifmmode~\acute{n}\else \'{n}\fi{}ski}, \citenamefont
  {Stagraczy\ifmmode~\acute{n}\else \'{n}\fi{}ski}, \citenamefont {Dugaev},\
  and\ \citenamefont {Chotorlishvili}}]{PhysRevB.106.224418}%
  \BibitemOpen
  \bibfield  {author} {\bibinfo {author} {\bibfnamefont {S.}~\bibnamefont
  {Wolski}}, \bibinfo {author} {\bibfnamefont {M.}~\bibnamefont {Inglot}},
  \bibinfo {author} {\bibfnamefont {C.}~\bibnamefont {Jasiukiewicz}}, \bibinfo
  {author} {\bibfnamefont {K.~A.}\ \bibnamefont {Kouzakov}}, \bibinfo {author}
  {\bibfnamefont {T.}~\bibnamefont {Mas\l{}owski}}, \bibinfo {author}
  {\bibfnamefont {T.}~\bibnamefont {Szczepa\ifmmode~\acute{n}\else
  \'{n}\fi{}ski}}, \bibinfo {author} {\bibfnamefont {S.}~\bibnamefont
  {Stagraczy\ifmmode~\acute{n}\else \'{n}\fi{}ski}}, \bibinfo {author}
  {\bibfnamefont {R.}~\bibnamefont {Stagraczy\ifmmode~\acute{n}\else
  \'{n}\fi{}ski}}, \bibinfo {author} {\bibfnamefont {V.~K.}\ \bibnamefont
  {Dugaev}},\ and\ \bibinfo {author} {\bibfnamefont {L.}~\bibnamefont
  {Chotorlishvili}},\ }\bibfield  {title} {\bibinfo {title} {Random spin-orbit
  gates in the system of a topological insulator and a quantum dot},\ }\href
  {https://doi.org/10.1103/PhysRevB.106.224418} {\bibfield  {journal} {\bibinfo
   {journal} {Phys. Rev. B}\ }\textbf {\bibinfo {volume} {106}},\ \bibinfo
  {pages} {224418} (\bibinfo {year} {2022})}\BibitemShut {NoStop}%
\bibitem [{\citenamefont {Khelashvili}\ and\ \citenamefont
  {Nadareishvili}(2022)}]{khelashvili2022generalized}%
  \BibitemOpen
  \bibfield  {author} {\bibinfo {author} {\bibfnamefont {A.}~\bibnamefont
  {Khelashvili}}\ and\ \bibinfo {author} {\bibfnamefont {T.}~\bibnamefont
  {Nadareishvili}},\ }\bibfield  {title} {\bibinfo {title} {Generalized
  heisenberg uncertainty relation in spherical coordinates},\ }\href@noop {}
  {\bibfield  {journal} {\bibinfo  {journal} {International Journal of Modern
  Physics B}\ }\textbf {\bibinfo {volume} {36}},\ \bibinfo {pages} {2250072}
  (\bibinfo {year} {2022})}\BibitemShut {NoStop}%
\bibitem [{\citenamefont {Judge}\ and\ \citenamefont
  {Lewis}(1963)}]{judge1963commutator}%
  \BibitemOpen
  \bibfield  {author} {\bibinfo {author} {\bibfnamefont {D.}~\bibnamefont
  {Judge}}\ and\ \bibinfo {author} {\bibfnamefont {J.}~\bibnamefont {Lewis}},\
  }\bibfield  {title} {\bibinfo {title} {On the commutator lz,/phi/@},\
  }\href@noop {} {\bibfield  {journal} {\bibinfo  {journal} {Phys. Letters}\
  }\textbf {\bibinfo {volume} {5}} (\bibinfo {year} {1963})}\BibitemShut
  {NoStop}%
\bibitem [{\citenamefont {Lombardi}(1980)}]{PhysRevA.22.797}%
  \BibitemOpen
  \bibfield  {author} {\bibinfo {author} {\bibfnamefont {J.~R.}\ \bibnamefont
  {Lombardi}},\ }\bibfield  {title} {\bibinfo {title} {Hydrogen atom in the
  momentum representation},\ }\href {https://doi.org/10.1103/PhysRevA.22.797}
  {\bibfield  {journal} {\bibinfo  {journal} {Phys. Rev. A}\ }\textbf {\bibinfo
  {volume} {22}},\ \bibinfo {pages} {797} (\bibinfo {year} {1980})}\BibitemShut
  {NoStop}%
\bibitem [{\citenamefont {Chotorlishvili}\ \emph {et~al.}(2017)\citenamefont
  {Chotorlishvili}, \citenamefont {Zi{\k{e}}ba}, \citenamefont {Tralle},\ and\
  \citenamefont {Ugulava}}]{chotorlishvili2017zitterbewegung}%
  \BibitemOpen
  \bibfield  {author} {\bibinfo {author} {\bibfnamefont {L.}~\bibnamefont
  {Chotorlishvili}}, \bibinfo {author} {\bibfnamefont {P.}~\bibnamefont
  {Zi{\k{e}}ba}}, \bibinfo {author} {\bibfnamefont {I.}~\bibnamefont
  {Tralle}},\ and\ \bibinfo {author} {\bibfnamefont {A.}~\bibnamefont
  {Ugulava}},\ }\bibfield  {title} {\bibinfo {title} {Zitterbewegung and
  symmetry switching in klein’s four-group},\ }\href@noop {} {\bibfield
  {journal} {\bibinfo  {journal} {Journal of Physics A: Mathematical and
  Theoretical}\ }\textbf {\bibinfo {volume} {51}},\ \bibinfo {pages} {035004}
  (\bibinfo {year} {2017})}\BibitemShut {NoStop}%
\end{thebibliography}%

\end{document}